\documentclass[aps,pra,reprint,twocolumn,floatfix,superscriptaddress,longbibliography]{revtex4-1}

\usepackage{tikz}
\usepackage{tikzscale}

\usepackage{pgfplots}
\pgfplotsset{compat=newest}
\usepgfplotslibrary{groupplots}
\usepgfplotslibrary{polar}
\usepgfplotslibrary{smithchart}
\usepgfplotslibrary{statistics}
\usepgfplotslibrary{dateplot}
\usetikzlibrary{arrows.meta}
\usetikzlibrary{backgrounds}
\usepgfplotslibrary{patchplots}
\usepgfplotslibrary{fillbetween}
\pgfplotsset{%
layers/standard/.define layer set={%
    background,axis background,axis grid,axis ticks,axis lines,axis tick labels,pre main,main,axis descriptions,axis foreground%
}{grid style= {/pgfplots/on layer=axis grid},%
    tick style= {/pgfplots/on layer=axis ticks},%
    axis line style= {/pgfplots/on layer=axis lines},%
    label style= {/pgfplots/on layer=axis descriptions},%
    legend style= {/pgfplots/on layer=axis descriptions},%
    title style= {/pgfplots/on layer=axis descriptions},%
    colorbar style= {/pgfplots/on layer=axis descriptions},%
    ticklabel style= {/pgfplots/on layer=axis tick labels},%
    axis background@ style={/pgfplots/on layer=axis background},%
    3d box foreground style={/pgfplots/on layer=axis foreground},%
    },
}

\usetikzlibrary{backgrounds}
\usetikzlibrary{arrows,calc,patterns}
\usetikzlibrary{shapes}
\usetikzlibrary{decorations.pathmorphing}
\usetikzlibrary{arrows.meta}
\usepackage{standalone}
\usepackage{amsfonts, amsmath, amssymb}
\usepackage{graphicx}
\usepackage{dcolumn}
\usepackage{bm}
\usepackage[normalem]{ulem}
\usepackage{color}
\usepackage[utf8]{inputenc}
\usepackage[colorlinks,citecolor=blue,linkcolor=blue,urlcolor=blue]{hyperref}
\usepackage{braket}
\usepackage[caption=false]{subfig}
\graphicspath{{img/}}
\usepackage{color}
\usepackage{soul}
\usepackage{mathtools}
\usepackage{float}
\newcommand{\gcor}{g_{1,2}^{(1)}}
\newcommand{\fcal}{\mathcal{F}}
\newcommand{\be}{\begin{equation}}
\newcommand{\ee}{\end{equation}}
\newcommand{\ba}{\begin{eqnarray}}
\newcommand{\ea}{\end{eqnarray}}
\newcommand{\bea}{\begin{equation}\begin{aligned}}
\newcommand{\eea}{\end{aligned}\end{equation}}

\newcommand{\gh}{\hat{\Gamma}}
\newcommand{\hh}{\hat{H}}

\newcommand{\lio}{\mathcal{L}}
\newcommand{\lioeff}{\mathcal{L}_\text{eff}}
\newcommand{\heff}{\hat{H}_\text{eff}}
\newcommand{\rhoss}{\hat{\rho}_{\text{SS}}}
\newcommand{\alphass}{\alpha_\text{SS}}

\newcommand{\aaa}{\hat{a}}

\newcommand{\bbb}{\hat{b}}
\newcommand{\dcal}{\mathcal{D}}

\newcommand{\ssx}{\hat{\sigma}^x}

\newcommand{\ssz}{\hat{\sigma}^z}

\newcommand{\sssz}{\hat{\sigma}^z}

\newcommand{\Jeff}{J_\text{eff}}
\newcommand{\Deff}{\Delta_\text{eff}}
\newcommand{\rhoh}{\hat{\rho}}

\def\rmi{{\rm {i}}}
\def\d{{\rm {d}}}
\def\tr{{\rm{Tr}}}

\include{physics}
\newcommand{\z}{\mathbb{Z}}

  {\left\lbrace\begin{array}{@{}l@{}}}%
  {\end{array}\right.}

\newcommand{\kett}{\rangle}
\newcommand{\braa}{\langle}

\def\ggg{{g_{1,2}^{(1)}}}

\begin{document}
\title{Dissipation-induced antiferromagnetic-like frustration in coupled photonic resonators}

\author{Zejian Li}
\affiliation{Universit\'{e} de Paris, Laboratoire Mat\'{e}riaux et Ph\'{e}nom\`{e}nes Quantiques, CNRS-UMR7162, 75013 Paris, France}

\author{Ariane Soret}
\affiliation{Universit\'{e} de Paris, Laboratoire Mat\'{e}riaux et Ph\'{e}nom\`{e}nes Quantiques, CNRS-UMR7162, 75013 Paris, France}
\affiliation{Complex Systems and Statistical Mechanics, Department of Physics and Materials Science, University of Luxembourg, L-1511 Luxembourg, Luxembourg}

\author{Cristiano Ciuti}
\affiliation{Universit\'{e} de Paris, Laboratoire Mat\'{e}riaux et Ph\'{e}nom\`{e}nes Quantiques, CNRS-UMR7162, 75013 Paris, France}

\begin{abstract}
We propose a photonic quantum simulator for anti-ferromagnetic spin systems based on reservoir engineering. We consider a scheme where quadratically driven dissipative Kerr cavities are indirectly coupled via lossy ancillary cavities. We  show that the ancillary cavities can produce an effective dissipative and Hamiltonian anti-ferromagnetic-like coupling between the cavities. By solving the master equation for a triangular cavity configuration, we demonstrate that the non-equilibrium steady state of the system bears full analogy with the ground state of an antiferromagnetic Ising model, exhibiting key signatures of frustration. We show that when the effective photon hopping amplitude is zero, the engineered non-local dissipation alone is capable of inducing antiferromagnetic interaction and frustration. This simple scheme can be generalised to arbitrary lattice geometries, providing a fully controllable recipe for simulating antiferromagnetism and frustration on a controlled quantum optical platform.
\end{abstract}

\date{\today}
\maketitle

\section{Introduction}
For decades, the physics of frustrated systems has gathered a great deal of interest as a fundamental problem in condensed matter physics. In a system with multiple constraints that cannot be satisfied simultaneously, the emerging frustration leads to interesting properties such as highly degenerate ground states \cite{ramirez1994strongly,moessner2006geometrical}, extensive entropy at zero temperature \cite{chalker1992hidden} and exotic phases of matter, with connections to high-$T_c$ superconductivity \cite{si2008strong,lehur09} or quantum critical phases \cite{ramires19}. Although at first studied in water ice \cite{giauque1936entropy}, the phenomenon of frustration has later been particularly explored in spin systems \cite{wannier1950antiferromagnetism,mezard1987spin,balents2010spin,zhou2017quantum,yan2011spin,shimizu2003spin,coldea2001experimental}, usually as a result of antiferromagnetic interaction combined with incompatible geometric constraints. A simple and paradigmatic model consists of antiferromagnetically interacting spins arranged on a triangular lattice, a system admitting a spin liquid phase as its ground state \cite{savary2016quantum}.

Recent impressive  developments in experimental techniques have triggered an increasing interest in the field of quantum simulation of spin systems using Rydberg atoms \cite{buchler10,sayrin20}, quantum gas microscopes \cite{kuhr16}, photonic simulators \cite{aspuru2012photonic,hartmann2016quantum,angelakis2017quantum} with semiconductors \cite{carusotto2013quantum,amo2016exciton,boulier2020microcavity,berloff2017realizing,goblot2019nonlinear} or circuit quantum electrodynamics (QED) \cite{haroche20,schmidt2013circuit,noauthor_abc_2020}. In particular,  driven-dissipative cavities subjected to two photon driving and dissipation have been explored both theoretically \cite{minganti2016exact,bartolo2016exact,rota2019quantum,bartolo2017homodyne,mirrahimi2014dynamically} and experimentally \cite{leghtas2015confining}. In such setups, a quadratic driving preserves the $\z_2$ parity symmetry of the photonic field and leads to a bimodal steady state - a mixture of coherent states with opposite phases, that can be mapped to spin states. These features make such setups not only suitable platforms for simulating quantum magnetism, but also a potential realisations of qubit for universal quantum computation \cite{mirrahimi2014dynamically}. Such cavities have been realised in circuit QED platforms \cite{leghtas2015confining,devoret2013superconducting,blais2007quantum} and can be engineered to be coupled to each other \cite{schoelkopf2008wiring}, allowing one to build artificial photonic lattices \cite{schmidt2013circuit,tsomokos2010using,houck2012chip}. However, their application on simulating frustrated spin systems is in its infancy. A recent theoretical study \cite{rota2019simulating} revealed that coupled quadratically-driven photonic cavities can simulate the antiferromagnetic Ising model \cite{wannier1950antiferromagnetism}, yet the model relies on negative photon hopping amplitude between cavities, the implementation of which remains a major challenge despite possible realisations with sophisticated techniques \cite{kounalakis2018tuneable,haddadi2014photonic}.

In this work, we propose a simple realisation of antiferromagnetic-like frustration in lattices of quadratically driven dissipative photonic cavities achieved via reservoir engineering. By indirectly coupling the target cavities (system) via lossy ancillary cavities (engineered reservoir), we obtain an effective description for the system with both antiferromagnetic-like Hamiltonian interaction (an effective photon hopping amplitude that can be tuned to be negative) and non-local dissipation coupling that is capable of inducing antiferromagnetic behavior in the system. By simulating the effective model via a consistently derived master equation for the reduced density matrix, we determine the first-order coherence correlation function and the Von Neumann entropy. We demonstrate that when applied to a triangular geometry, our scheme yields a simulator for antiferromagnetically coupled Ising spins exhibiting key signatures of frustration.

This article is structured as follows. In Sec. \ref{sec:setup} we present the considered system consisting of target and ancillary cavities and then derive the effective dynamics for the target cavities.  In Sec. \ref{sec:res} we present and discuss numerical results for the triangular geometry. Finally, we draw our conclusions and perspectives in Sec. \ref{sec:end}.

\section{System and theoretical model}\label{sec:setup}
Let us consider a 1D chain of $N$ pairs of single-mode cavities with annihilation operators $\{\hat{a}_1,\hat{b}_1,\hat{a}_2,\hat{b}_2,\cdots,\hat{a}_N,\hat{b}_N\}$ and periodic boundary conditions.
The target cavities are described by the bosonic mode annihilation operators $\hat{a}_j$ while the lossy reservoir cavities by the operators  $\hat{b}_j$. 
\begin{figure}[t!]
    \centering
	\includegraphics[width=\linewidth]{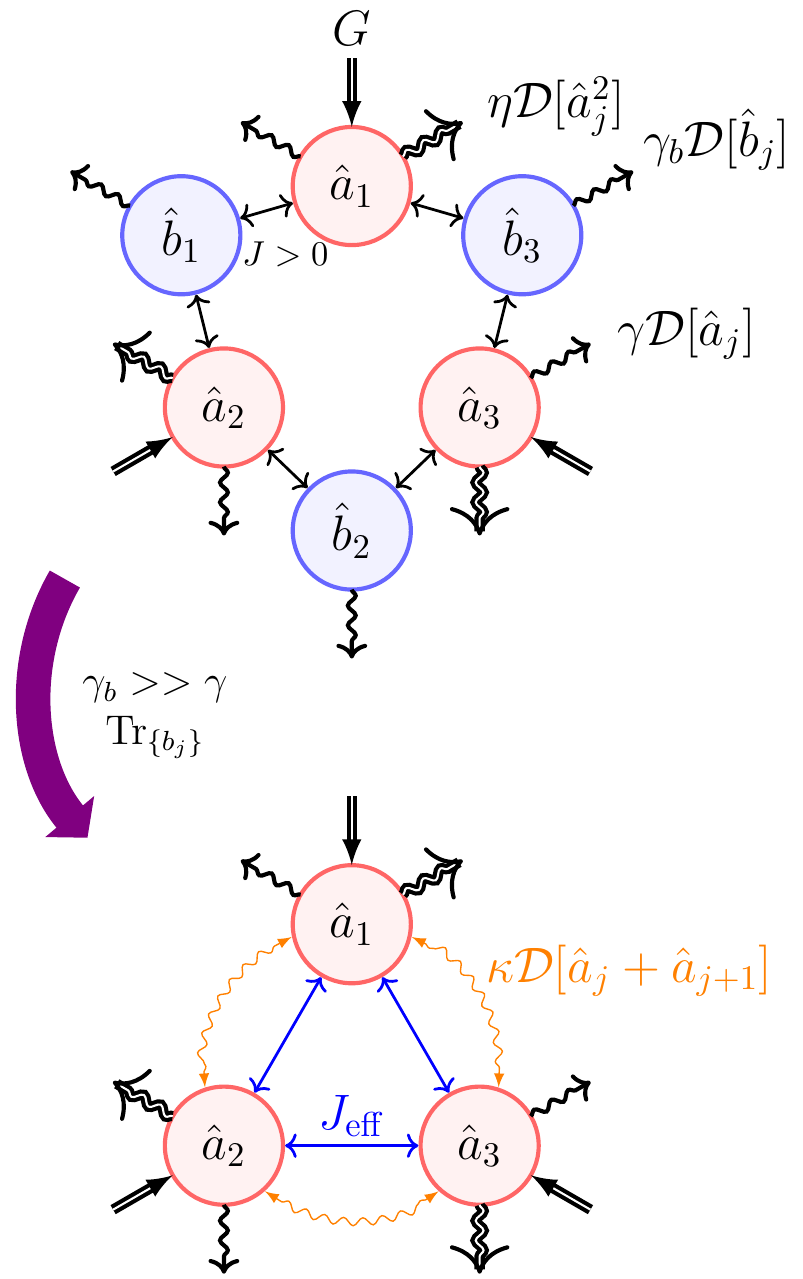}
    \caption{Schematic of the considered system for the case with $N=3$ target cavities, where $\hat{a}_j$ is the photon annihilation operator on the $j$-th cavity, $G$ is the two-photon driving amplitude,   $\gamma$ is the single-photon loss rate, and $\eta$ is the two-photon loss rate. The target cavities are coupled indirectly via the undriven lossy ancillary cavities: $\hat{b}_j$ is the corresponding ancillary mode annihilation operator and  $\gamma_b\gg\gamma$ is its single-photon loss. The hopping coupling constant $J$ between ancilla and target cavities is assumed to be positive. The effective model for the target cavities is obtained by tracing out the ancillary degrees of freedom. This produces an effective coupling between  target cavities that has both a coherent contribution (via the photon hopping  $\Jeff$) and a dissipative part (via the dissipator $\kappa\dcal[\aaa_j+\aaa_{j+1}]$). The effective hopping amplitude $\Jeff$ can be tuned to be negative, positive or zero depending on the choice of parameters. The nonlocal dissipator has a symmetric jump operator that favors antiferromagnetic-like correlations.}
    \label{fig:3cav}
\end{figure}
Each target cavity is coupled to the neighboring reservoir cavities via the hopping coupling with amplitude $J~(>0)$. Each target cavity is assumed to have a mode frequency $\omega_0$ and Kerr nonlinearity $U$ and subjected to a coherent two-photon drive with amplitude $G$, driving frequency $\omega_p$ and two-photon dissipation rate $\eta$. The ancillary cavity modes have frequency $\omega_0-\delta\-\omega$ and are assumed to be undriven and linear. We further assume the presence of single-photon loss for both the target sites (with rate $\gamma$) and the reservoir cavities (with rate $\gamma_b$). The considered system is schematically depicted in Fig.\ref{fig:3cav} for the case $N=3$.

In a frame rotating at the frequency $\omega_p/2$, the   Hamiltonian of the considered system reads ($\hbar=1$):
\bea
        \hat{H}=& \sum_j\hat{H}_j \, ,\\
    \hat{H}_j=&-\Delta\hat{a}_j^\dag\hat{a}_j-(\Delta+\delta\omega)\hat{b}_j^\dag\hat{b}_j\\&-J[(\hat{a}_j+\hat{a}_{j+1})\hat{b}_j^\dag+(\hat{a}_j^\dag+\hat{a}_{j+1}^\dag)\hat{b}_j]\\
    &+ \dfrac{U}{2}\hat{a}_j^{\dag2}\hat{a}_j^2+ \dfrac{G}{2}\hat{a}_j^{\dag2}+\dfrac{G^\ast}{2}\hat{a}_j^2.
\eea
where $\Delta=\omega_p/2-\omega_0$ is the pump-cavity detuning. Under the Born-Markov approximation, the system can be described by the density matrix $\hat{\rho}$ whose dynamics is governed by the Lindblad master equation $\dfrac{\d\rhoh}{\d t}= {\mathcal L} \rhoh$ where ${\mathcal L}$ is the Liouville operator, defined as follows:
\bea\label{eq:meq}
\dfrac{\d\rhoh}{\d t}=&-\rmi[\hh,\rhoh]+\sum_j\big(\gamma \mathcal{D}[\hat{a}_j]+  \gamma_b\mathcal{D}[\hat{b}_j]+\eta\mathcal{D}[\hat{a}^2_j]\big)\rhoh,
\eea
where the action of the dissipator $\dcal[\gh]$ on the density matrix is defined as 
\bea
\dcal[\gh]\rhoh=\gh\rhoh\gh^\dag-\dfrac{1}{2}\{\gh^\dag\gh,\rhoh\} \, ,
\eea
where $\gh$ is the so-called jump operator. We are interested in the regime where $\gamma_b\gg\gamma$, such that the modes $\hat{b}_j$ can be traced out with standard adiabatic elimination techniques. Defining
\begin{align}
        \Jeff =& \dfrac{-J^2(\Delta+\delta\omega)}{\frac{\gamma_b^2}{4}+(\Delta+\delta\omega)^2},\\
        \Delta_\text{eff} =& z \Jeff + \Delta,\\
        \kappa=&\gamma_b|g|^2=\dfrac{\gamma_b J^2}{\dfrac{\gamma_b^2}{4}+(\delta\omega+\Delta)^2},
\end{align}
where $z=2$ (the coordination number) in the case of a 1D chain,
we can write the effective Hamiltonian for the target sites as
\begin{align}\label{eq:heff}
    \hat{H}_{\text{eff}}=& -\Delta_\text{eff}\sum_j\hat{a}_j^\dag\hat{a}_j - \Jeff\sum_{<j,j'>}(\hat{a}_j^\dag\hat{a}_{j'}+\hat{a}_j\hat{a}_{j'}^\dag)\\
     &+ \dfrac{U}{2}\hat{a}_j^{\dag2}\hat{a}_j^2+ \dfrac{G}{2}\hat{a}_j^{\dag2}+\dfrac{G^\ast}{2}\hat{a}_j^2,
\end{align}
with the dissipators $\gamma \mathcal{D}[\hat{a}_j]$, $\eta\dcal[\aaa^2_j]$ and $\kappa\mathcal{D}[\hat{a}_j+\hat{a}_{j+1}]$.
From now on, it will be convenient to work with the effective parameters $\Delta_\text{eff}$, $\Jeff$ and $\kappa$ as the original Hamiltonian parameters can be obtained as functions of them:
\begin{align}
    \Delta =& \Delta_\text{eff} - z \Jeff,\\
    \delta\omega =& -\dfrac{\gamma_b \Jeff}{\kappa} - (\Delta_\text{eff} - z \Jeff),\\
    J =& \sqrt{\dfrac{\kappa\gamma_b}{4}+\dfrac{\gamma_b \Jeff^2}{\kappa}},
\end{align}
and the transformation is well-defined as long as $\kappa>0$. The effective master equation for the reduced density matrix of the target system is a Lindlad equation described by the effective Liouvillian as $\lioeff$, defined as 
\bea
\lioeff(\cdot)=&-\rmi[\heff,\cdot]\\&+\sum_j(\gamma\dcal[\aaa_j]+\eta\dcal[\aaa^2_j]+\kappa\dcal[\aaa_j+\aaa_{j+1}]).
\eea

At this stage, it is already important to point out that:
\begin{enumerate}
    \item The nearest neighbours in the effective model are  dissipatively coupled via the dissipators $\kappa\dcal[\aaa_j+\aaa_{j+1}]$, that preserves the $\z_2$ symmetry of the system (invariance under a global sign change $\aaa_j\rightarrow-\aaa_j~\forall j$) and are capable of inducing frustration, as we will show in the next section.
    \item The effective photon hopping amplitude $\Jeff$ can be tuned and can be also negative when $\Delta+\delta\omega>0$.
\end{enumerate}

As studied in Ref.\cite{rota2019simulating}, in the limit of $G/\gamma\rightarrow\infty$, each cavity will be driven into a statistical mixture of two coherent states with opposite phase $\ket{\pm\alpha}$. Indeed, the steady state can be mapped to Ising spins with the identification \footnote{In Ref. \cite{rota2019quantum}, the photonic Hamiltonian was actually mapped into a Heisenberg $XY$ Hamiltonian by identifying the Schroedinger cat states $\propto (\vert \alpha \rangle \pm \vert - \alpha \rangle )$ with the spin states along the $z$-direction. However, in the regime of $\vert \alpha \vert \gg 1$, only the $\ssx_j\ssx_{j+1}$ terms survive in the $XY$ Hamiltonian and $\ket{\pm\alpha}$ are asymptotically the $\ssx$ eigenstates. Therefore, we can identify it with an Ising model after a rotation of basis $\ssx\rightarrow\ssz$.} $\ket{\alpha}\rightarrow\ket{\uparrow},~\ket{-\alpha}\rightarrow\ket{\downarrow}$, since we have 
\bea
\lim_{\vert \alpha \vert \rightarrow\infty}\braa-\alpha|\alpha\kett=\lim_{\vert \alpha \vert \rightarrow\infty}\exp{(-2|\alpha|^2)}=0.
\eea
 The operator $\aaa_j$ can be mapped to $\alpha\sssz_j$ when projected onto the spin basis 
in the limit of large driving. Therefore, from the spin point of view, the Hamiltonian (\ref{eq:heff}) gives an effective Ising interaction $\sssz_j\sssz_{j'}$ with coupling constant proportional to $\Jeff$. 
The non-local dissipator $\dcal[\aaa_j+\aaa_{j+1}]$ is expected to induce anti-alignment of nearest neighbours, i.e.  $\ket{\pm\alpha,\mp\alpha,\pm\alpha,\mp\alpha,\cdots}$. In fact, the jump operator destroys excitations where there is alignment.

To investigate the behavior of the steady state of the system, we numerically solve the master equation using the effective model to obtain the steady state density matrix $\rhoss$ that satisfies
$ \lioeff\rhoss=0 $, 
where the detuning is set to $\Deff=\Jeff$ in order to favor the $k=\pi$ modulation of the photonic field $\ket{\pm\alpha,\mp\alpha,\pm\alpha,\mp\alpha,\cdots}$ \cite{rota2019simulating} (the phase of the driven cavity field changes by $\pi$ moving from one cavity to the nearest one). 
We will be interested in the first-order coherence correlation function, defined as 
\begin{equation}
    g_{1,2}^{(1)}=\dfrac{\text{Tr}[\rhoss\hat{a}_1^\dag\hat{a}_2]}{\text{Tr}[\rhoss\hat{a}_1^\dag\hat{a}_1]},
\end{equation}
and the Von Neumann entropy 
\begin{equation}
    S = -\text{Tr}[\rhoss\ln{\rhoss}].
\end{equation}
Note that with the mapping $\aaa_j\rightarrow\alpha\sssz_j$, we have
$
\ggg\simeq\braa \sssz_1\sssz_2\kett
$
for $\vert \alpha \vert \gg 1$, i.e. for sufficiently strong driving.

\begin{figure}
    \centering
    \includegraphics[width=\linewidth]{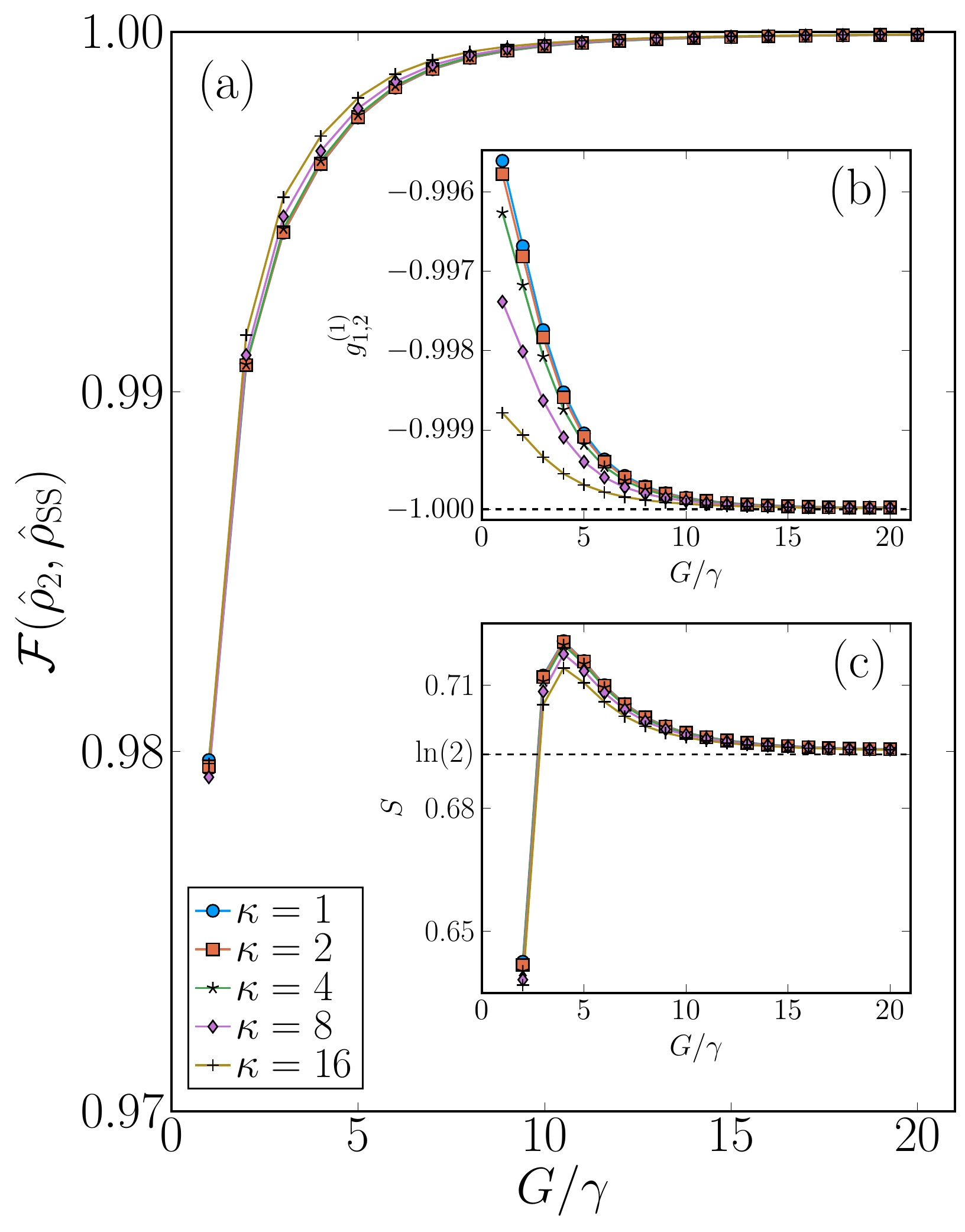}
    \caption{Steady-state behavior of the $N=2$ system with $\Deff=\Jeff=-5\gamma,~U=4\gamma$ and $\eta=\gamma$. (a) The fidelity $\fcal$ between the numerical solution $\rhoss$ and the ansatz $\rhoh_2$ is plotted versus the two-photon driving amplitude $G$. (b) and (c) report the corresponding values of the first-order coherence correlation function $\ggg$ and of the Von Neumann entropy $S$. Different points (see legend) correspond to different values of the effective nonlocal dissipation rate $\kappa$.}
    \label{fig:2cav_all}
\end{figure}

\section{Results and discussion}\label{sec:res}

To reveal the antiferromagnetic behaviour of the considered system, we first investigate the case with $N=2$ sites. In this dimer configuration, we expect to see the antiferromagnetic ordering since there is no geometric frustration. In Fig. \ref{fig:2cav_all} we present the results for a finite value of the effective photon hopping amplitude $\Jeff=-5\gamma<0$ and different values of the nonlocal dissipative coupling $\kappa$. As the driving $G$ increases, the correlation $\ggg$ converges to $-1$, directly witnessing the antiferromagnetic alignment of the simulated spins in the two sites. Moreover, the entropy converges to $\ln(2)$ for all values of $\kappa$. This suggests that the steady state density matrix can be approximated by the ansatz
\bea\label{eq:rho2}
    \rhoh_2(\alpha)=\dfrac{1}{2}(|\alpha,-\alpha\rangle\langle\alpha,-\alpha| +|-\alpha,\alpha\rangle\langle-\alpha,\alpha|)
\eea
in the strong driving limit. Indeed, as shown in the figure, the fidelity $\fcal$ between the steady-state density matrix $\rhoss$ and the ansatz $\rhoh_2(\alphass)$ converges to $1$ for increasing driving $G$. Such fidelity is defined as
\be
    \mathcal{F}\left(\rhoss,\rhoh_2(\alphass)\right)=\left\lvert\tr\left(\sqrt{\sqrt{\rhoh_2}\rhoss\sqrt{\rhoh}_2}\right)\right\lvert^2,
\ee
where $\alphass=\sqrt{\tr(\rhoss\aaa_1^2)}$. 

\begin{figure}
    \centering
    \includegraphics[width=\linewidth]{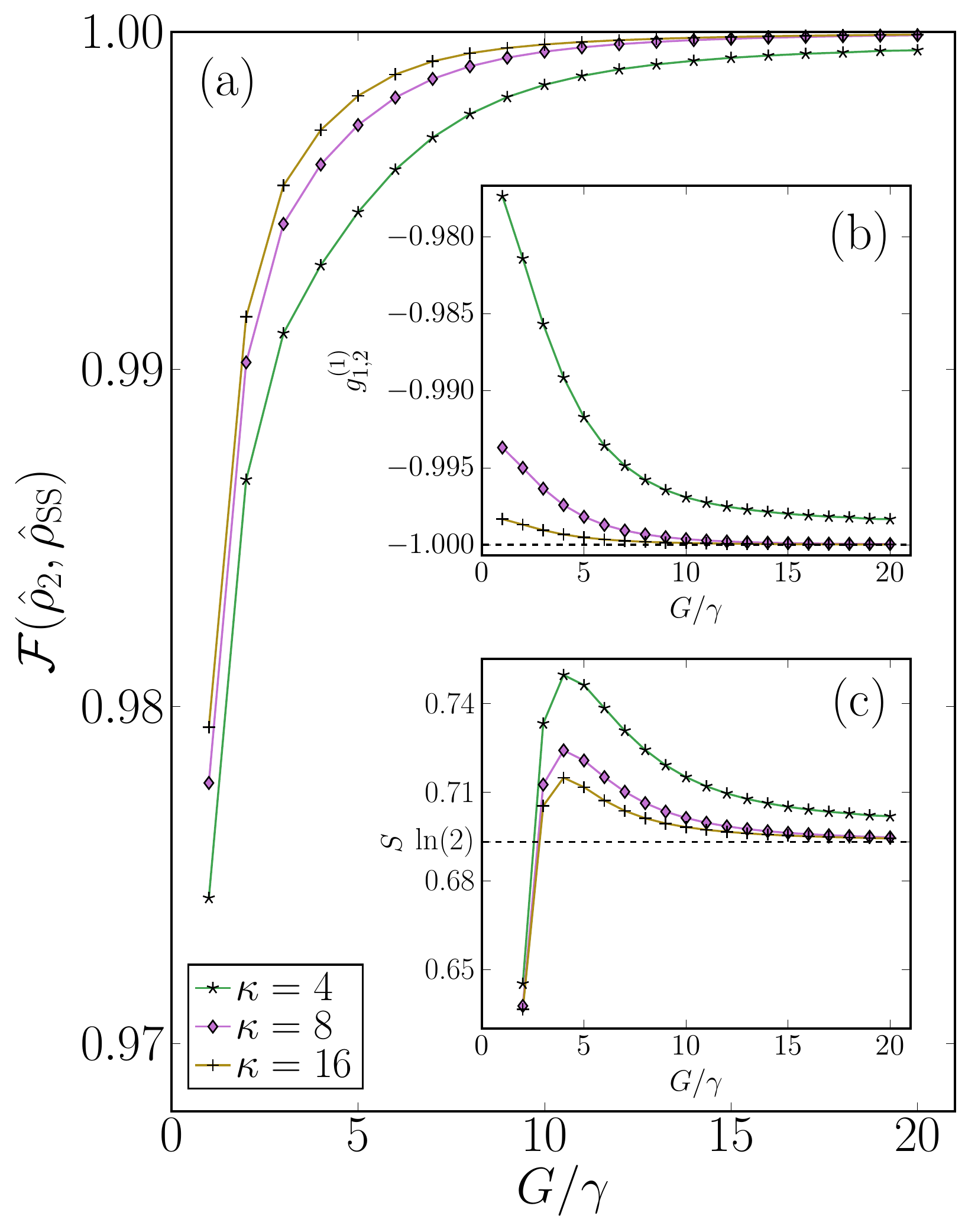}
    \caption{Same as Fig. \ref{fig:2cav_all}, but with the parameters $\Deff=\Jeff=0,~U=4\gamma$ and $\eta=\gamma$. The coupling here is purely dissipative and displays the key antiferromagnetic signatures.}
    \label{fig:2cav_diss_all}
\end{figure}

Note that when the dissipative coupling strength $\kappa$ increases, we achieve also a faster convergence, which implies that the dissipator $\kappa\dcal[\aaa_j+\aaa_{j+1}]$ enhances the antiferromagnetic interaction. Importantly, the nonlocal dissipative coupling {\it alone} is sufficient to obtain the key antiferromagnetic signatures (i.e.  $\ggg \to 1$, $S \to \ln(2)$ and $\fcal \to 1$), as shown in Fig.\ref{fig:2cav_diss_all} where  $\Jeff=0$.

We now consider the more interesting case of $N=3$ where geometric frustration can emerge. Similar to the $N=2$ case, we expect the steady state density matrix to be approximated by the ansatz
\begin{equation}\label{eq:rho3}
    \begin{aligned}
    \rhoh_3(\alpha)=&\dfrac{1}{6}(|\alpha,\alpha,-\alpha\rangle\langle\alpha,\alpha,-\alpha|\\
    &+|\alpha,-\alpha,\alpha\rangle\langle\alpha,-\alpha,\alpha|\\
    &+|-\alpha,\alpha,\alpha\rangle\langle-\alpha,\alpha,\alpha|\\
    &+|-\alpha,-\alpha,\alpha\rangle\langle-\alpha,-\alpha,\alpha|\\
    &+|-\alpha,\alpha,-\alpha\rangle\langle-\alpha,\alpha,-\alpha|\\
    &+|\alpha,-\alpha,-\alpha\rangle\langle\alpha,-\alpha,-\alpha|),
    \end{aligned}
\end{equation}
where we have a clear analogy with the six-fold degenerate ground state of the antiferromagnetic triangular Ising model. 

\begin{figure}[t!]
    \centering
    \includegraphics[width=0.9\linewidth]{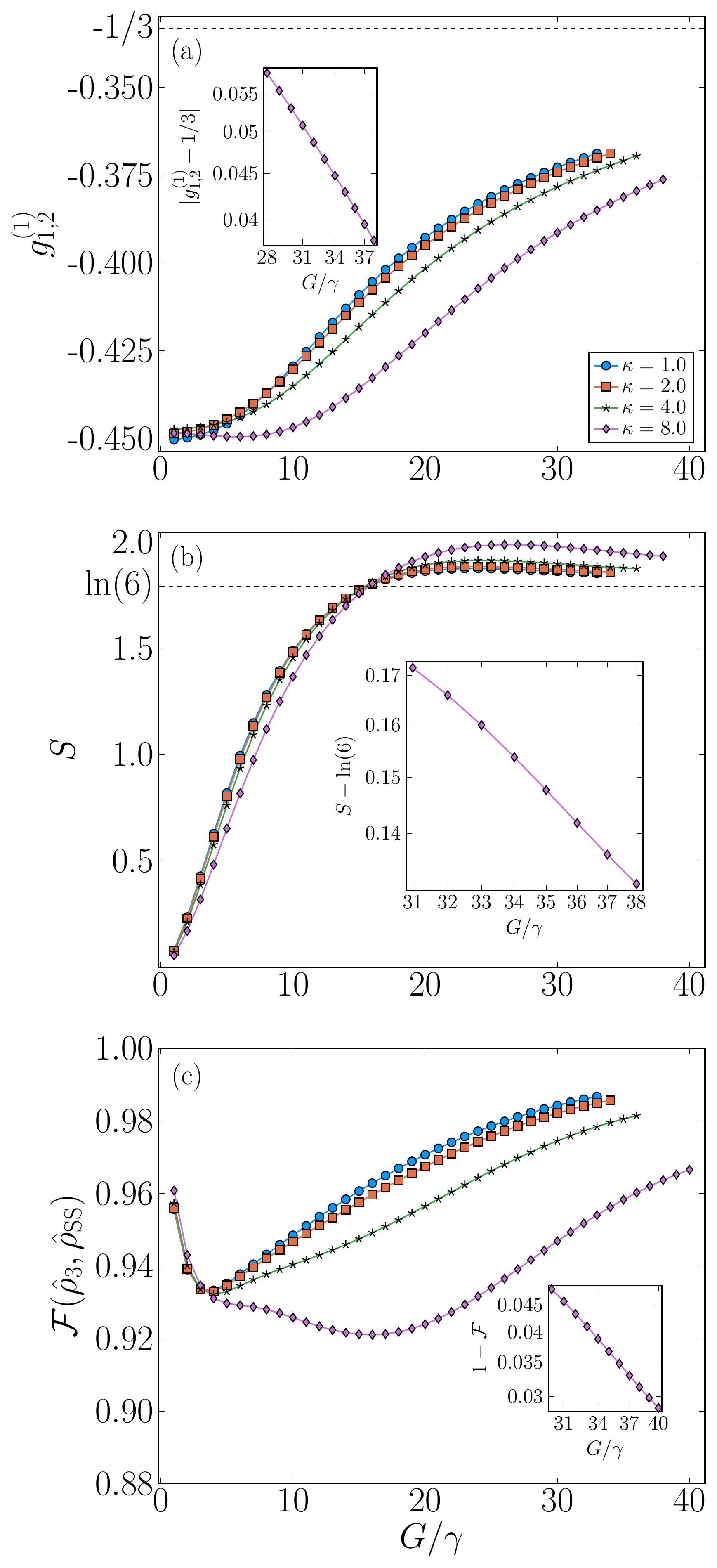}
    \caption{Steady-state behavior of the $N=3$ triangular system for $\Deff=\Jeff=-5\gamma,~U=10\gamma$ and $\eta=\gamma$. All quantities are plotted versus the two-photon driving $G$. The panels reports plots of (a) the first-order coherence correlation function $\gcor$ (inset: the quantity $|\gcor+1/3|$), (b) the Von Neumann entropy $S$ (inset: the quantity $S-\ln(6)$) and (c) the fidelity $\fcal$ between the numerical solution $\rhoss$ and the ansatz $\rhoh_3$ (inset: the quantity 1-$\fcal$). The insets are all plotted in $\log$-$\log$ scale, showing the asymptotic convergence of the respective quantities.}
    \label{fig:tri_all}
\end{figure}

We first demonstrate that with a finite value of $\Jeff<0$, our model is capable of simulating the frustrated Ising spins. Fig. \ref{fig:tri_all} summarizes the steady-state behavior of our model as a function of driving $G$ for different values of the nonlocal dissipation rate $\kappa$. 
As the driving increases, the value of the first-order coherence correlation function $\ggg$ converges asymptotically to $-1/3$, which is also the spin correlation value in the corresponding antiferromagnetic triangular Ising model \cite{stephenson1964ising}. The Von Neumann entropy $S$ converges asymptotically to $\ln{(6)}$, agreeing with the six-fold degenerate ground state of the simulated antiferromagnetic Ising mode. Furthermore, the fidelity $\fcal$ of the density matrix $\rhoss$ with respect to the ansatz $\rhoh_3$ also converges to $1$, validating the analogy with the spin system we made previously.

\begin{figure}
    \centering
    \includegraphics[width=0.9\linewidth]{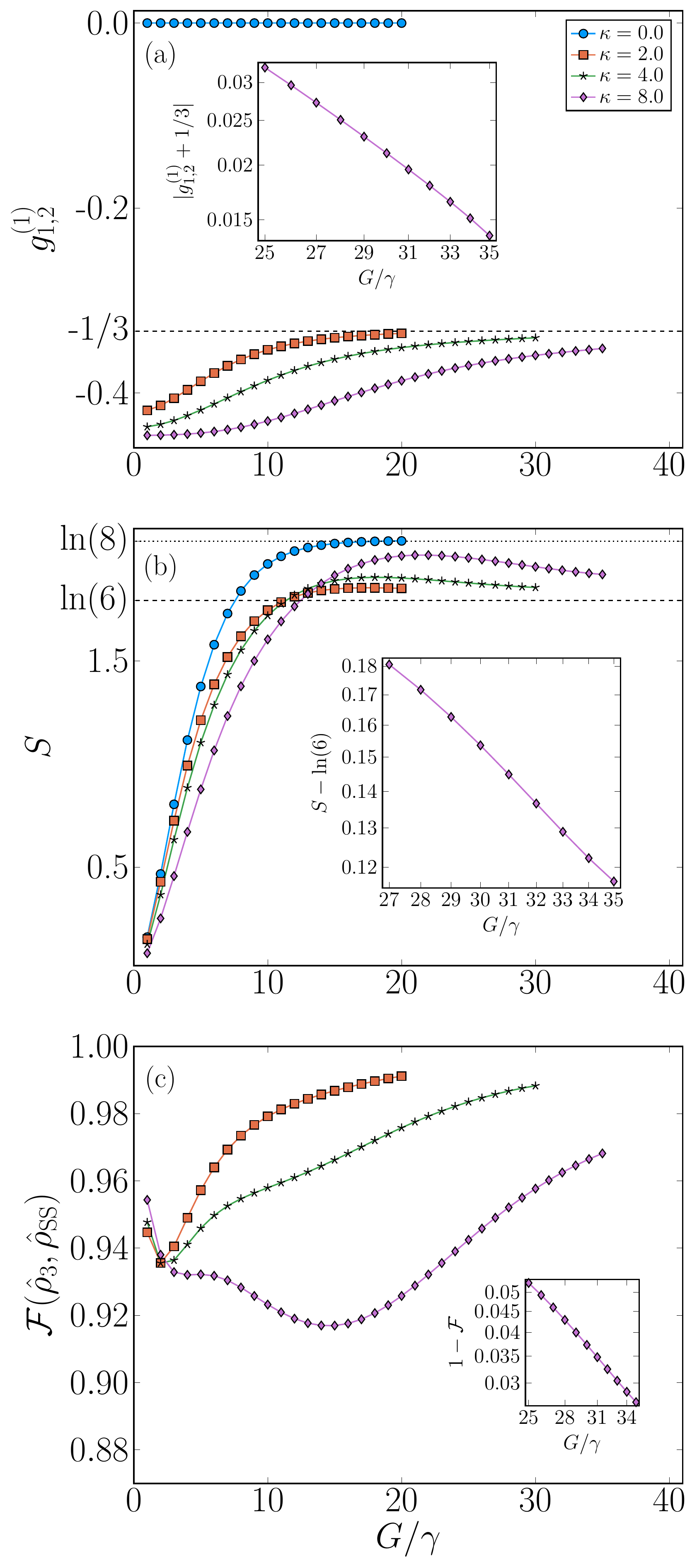}
    \caption{Same as Fig. \ref{fig:tri_all} but with $\Deff=\Jeff=0,~U=10\gamma$ and $\eta=\gamma$. Note that here the antiferromagnetic frustration effects are purely of dissipative nature via the nonlocal dissipative coupling.}
    \label{fig:tri_all_diss}
\end{figure}

Our most important result is for the case of $\Delta=\Jeff=0$ and $\kappa>0$, as summarised in Fig. \ref{fig:tri_all_diss}. Despite the absence of coherent antiferromagnetic interaction in the Hamiltonian, we successfully recovered the key signatures of frustration ($\ggg \to -1/3$, the entropy $S \to \ln(6)$ and  $\fcal(\rhoh_3,\rhoss) \to 1$. For comparison, we also simulated the trivial hypothetical scenario of $\kappa=0$ \footnote{Note that in our effective model we always have $\kappa>0$. The case where $\kappa=0$ is simulated only for illustrative purpose to show the direct effect of the dissipative coupling.}, in which case the correlation $\ggg = 0$ as the modes $\aaa_j$ are entirely decoupled, and the entropy tends to $\ln(8)$ instead of $\ln(6)$, corresponding to the $2^3=8$ fold degeneracy of the ground state of the non-interacting triangular  model. This highlights the fact that the frustration in the case of $\Jeff=0$ is directly induced by the dissipative coupling $\kappa\mathcal{D}[\hat{a}_j+\hat{a}_{j+1}]$.

\section{Conclusion and outlook}\label{sec:end}
In this work, we have proposed a reservoir engineering scheme allowing for the quantum simulation of frustrated Ising antiferromagnets with coupled photonic resonators subjected to coherent two-photon pumping. We have shown theoretically that the proposed configuration displays a dissipative coupling inducing antiferromagnetic-like behavior and frustration even when the effective photon hopping amplitude is zero.
By numerically solving the master equation for the cases with two and three sites respectively, we demonstrated the full analogy between the steady-state of our model and the antiferromagnetic Ising model supported by the first-order coherence correlation and the Von Neumann entropy. 

The scheme proposed here provides a building block for simulating antiferromagnetic spin lattices of arbitrary geometry, where the interaction depends on the easily tunable coherent photon hopping amplitude and the dissipative coupling rate, and which can be  implemented in photonic platforms. 

\acknowledgements{We would like to acknowledge  discussions with Z. Denis as well as for numerical help. This work was supported by the FET FLAGSHIP Project PhoQuS (grant agreement ID: 820392) and by Project NOMOS (ANR-18-CE24-0026).}

\appendix
\section{Adiabatic elimination of \texorpdfstring{mode $\bbb_j$}{the reservoir modes}}
As the modes described by the bosonic operators $\hat{b}_j$ have no direct coupling between them, we can eliminate them independently. In order to eliminate the mode $\hat{b}_j$ for a given $j\in\{1,\dots,N\}$, we start by regrouping the Liouvillian superoperator into terms involving $\hat{b}_j$ and all the rest, in such a way that the master equation of the total density of the system $\hat{\rho}$ can be written as
\begin{align}\label{eq:tot}
    \dfrac{\d }{\d t}\hat{\rho}=&\mathcal{L}_{\hat{b}_j}\hat{\rho}+\mathcal{L}_s\hat{\rho},
\end{align}
where
\begin{align}
    \mathcal{L}_{\hat{b}_j}(\cdot)=&-i[\hat{H}_{\hat{b}_j},\cdot]+\gamma_b\mathcal{D}[\hat{b}_j](\cdot),\\
    \hat{H}_{b_j}=&-(\Delta+\delta\omega)\hat{b}_j^\dag\hat{b}_j\\&-J[(\hat{a}_j+\hat{a}_{j+1})\hat{b}_j^\dag+(\hat{a}_j^\dag+\hat{a}_{j+1}^\dag)\hat{b}_j],\nonumber\\
    \mathcal{D}[\hat{b}_j](\cdot)=&\hat{b}_j(\cdot)\hat{b}_j^\dag-\dfrac{1}{2}\big\{\hat{b}^\dag_j\hat{b}_j,\cdot\big\}
\end{align}
with $\mathcal{L}_s$ including all the terms not depending on $\hat{b}_j$. As the mode $\hat{b}_j$ is strongly dissipated with no direct pumping, we assume that it always stays close to the vacuum state. Hence, we can develop the full density matrix as 
\bea
    \hat{\rho}=&\hat{\rho}_{00}|0\rangle\langle0|+\delta(\hat{\rho}_{01}|0\kett\braa1|+\hat{\rho}_{10}|1\kett\braa0|)\\&+\delta^2(\hat{\rho}_{11}|1\kett\braa1|+\hat{\rho}_{02}|0\kett\braa2|+\hat{\rho}_{20}|2\kett\braa0|)\\&+O(\delta^3),
\eea
\newcommand{\Tr}{\text{Tr}}
where $|m\kett\braa n|$ acts on the Hilbert space of mode $\hat{b}_j$, $\hat{\rho}_{mn}$ acts on the Hilbert space corresponding to the rest of the system and $\delta$ is a small parameter. Furthermore, we assume $\gamma/\gamma_b\sim\delta^2$ and $G/\gamma_b\sim U/\gamma_b\sim J/\gamma_b\sim\delta$. We aim to find the effective dynamics of the reduced density matrix $\hat{\rho}_s=\Tr_{\hat{b}_j}[\hat{\rho}]=\hat{\rho}_{00}+\delta^2\hat{\rho}_{11}$ up to second order in $\delta$ where the ancillary mode $\hat{b}_j$ is traced out. First note that
\begin{align}
    \dfrac{1}{\gamma_b}\dfrac{\d}{\d t}\hat{\rho}_{00}=&\dfrac{1}{\gamma_b}\mathcal{L}_s(\hat{\rho}_{00})-\rmi\delta^2(\hat{A}^\dag\hat{\rho}_{10}-\hat{\rho}_{01}\hat{A})\\&+\delta^2\hat{\rho}_{11}+O(\delta^3)\nonumber,\label{eq:rho00}\\
\dfrac{1}{\gamma_b}\dfrac{\d}{\d t}\hat{\rho}_{10}=&-\rmi\hat{A}\hat{\rho}_{00}+\big(\rmi\dfrac{\Delta+\delta\omega}{\gamma_b}-\dfrac{1}{2})\hat{\rho}_{10}+O(\delta),\\
\dfrac{1}{\gamma_b}\dfrac{\d}{\d t}\hat{\rho}_{11}=&-\rmi(\hat{A}\hat{\rho}_{01}-\hat{\rho}_{10}\hat{A})-\hat{\rho}_{11}+O(\delta),\\
\hat{\rho}_{01}=&\hat{\rho}_{10}^\dag,
\end{align}
where we define $\hat{A}=-\dfrac{J}{\delta\gamma_b}(\hat{a}_j+\hat{a}_{j+1})$. With the adiabatic assumption, we can approximate that $\hat{\rho}_{10}$, $\hat{\rho}_{01}$ and $\hat{\rho}_{11}$ are constantly in their steady values on time scales much larger than $\gamma_b^{-1}$. This gives 
\begin{align}
    \hat{\rho}_{10}=&\dfrac{\rmi \gamma_b}{\rmi(\Delta+\delta\omega)-\frac{\gamma_b}{2}}\hat{A}\hat{\rho}_{00}+O(\delta),\\
    \hat{\rho}_{11}=&\dfrac{\gamma_b^2}{(\Delta+\delta\omega)^2+\frac{\gamma_b^2}{4}}\hat{A}\hat{\rho}_{00}\hat{A}^\dag+O(\delta).
\end{align}
Inserting these terms back into Eq.(\ref{eq:rho00}) we obtain the master equation for the reduced density matrix:
\begin{align}
    \dfrac{\d}{\d t}\hat{\rho}_s=&\mathcal{L}_s(\hat{\rho}_s)+\mathcal{L}_{\text{eff},j}(\hat{\rho}_s),\\
    \mathcal{L}_{\text{eff},j}=&-\rmi[\hat{H}_{\text{eff},j},\cdot]+\kappa\mathcal{D}[\hat{a}_j+\hat{a}_{j+1}],\\
    \hat{H}_{\text{eff},j}=&-J_\text{eff}(\hat{a}_j^\dag\hat{a}_{j+1}+\hat{a}_j\hat{a}_{j+1}^\dag)\\&\nonumber-J_\text{eff}(\hat{a}_j^\dag\hat{a}_j+\hat{a}_{j+1}^\dag\hat{a}_{j+1}),\\
    J_\text{eff}=&-\dfrac{J^2(\Delta+\delta\omega)}{\frac{\gamma_b^2}{4}+(\Delta+\delta\omega)^2},\\
    \kappa=&\dfrac{\gamma_b J^2}{\frac{\gamma_b^2}{4}+(\Delta+\delta\omega)^2}.
\end{align}
This results in a coupling between $\hat{a}_j$ and $\hat{a}_{j+1}$ that has both an Hamiltonian part proportional to the effective hopping rate $J_\text{eff}$ and a nonlocal dissipative interaction proportional to $\kappa$. Eliminating all the ancillary modes $\hat{b}_j$ gives the effective model that has been presented in the main text of the paper.

\section{Benchmarking of the effective model against exact results}
To benchmark the effective model we derived above, we simulate the $N=2$ system using the full master equation and compare the results with those obtained using the effective model. Note that as we have only two target sites, it suffices to consider only one ancilla cavity ($b_1$), sandwiched between the two targets ($a_1,a_2$), in the full simulation. We denote the steady-state density matrix of the full model by $\rhoss^\mathrm{full}$, obtained by solving the master equation (Eq. (\ref{eq:meq}) in the main text):
\bea
\lio\rhoss^\mathrm{full}=0.
\eea
Tracing out the ancilla mode gives the reduced density matrix $\rhoss^\mathrm{full,r}$ for the system represented by the target modes:
\bea
\rhoss^\mathrm{full,r}=\Tr_{\bbb_1}[\rhoss^\mathrm{full}].
\eea
We denote also the steady-state density matrix of the effective model by $\rhoss$ (as in the main text), which is determined by
\bea
\lioeff\rhoss=0.
\eea
To quantify the benchmarking, we have calculated the fidelity $\fcal$ between the two solutions, defined as $\fcal = \fcal(\rhoss,\rhoss^\mathrm{full,r})
$, of course by using the same system parameters.
To demonstrate the validity of the effective model, here we report results of simulations for $\Deff=\Jeff=-5\gamma,~U=4\gamma$ and $\eta=\gamma$, which are the same parameters used to calculate Fig. \ref{fig:2cav_all} in the main text, using different values of $\gamma_b/\gamma$. 
\begin{figure}
    \centering
    \includegraphics[width=\linewidth]{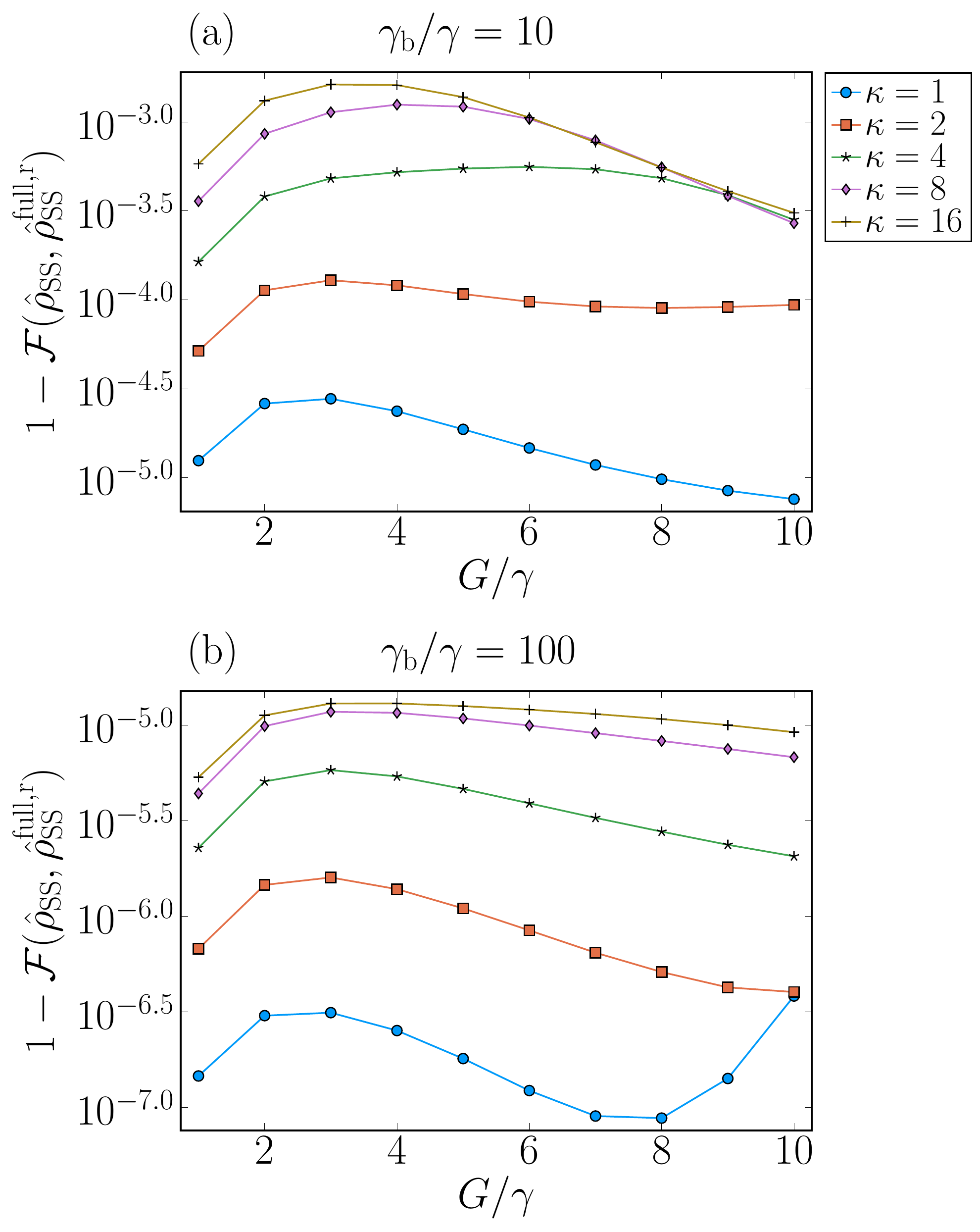}
    \caption{The infidelity $1-\fcal$ of the effective model steady-state density matrix $\rhoss$ with respect to the reduced density matrix $\rhoss^\mathrm{full,r}$ calculated from the full solution as a function of the driving $G$ for different values of the nonlocal dissipative coupling $\kappa$. Dissipation parameters: (a) $\gamma_b/\gamma=10$ and (b) $\gamma_b/\gamma=100$. The other parameters are $\Deff=\Jeff=-5\gamma$, $U=4\gamma$ and $\eta=\gamma$. The effective model is extremely accurate in a wide range of parameters as witnessed by the very small infidelities.}
    \label{fig:2cav_bm}
\end{figure}
As shown in Fig. \ref{fig:2cav_bm}, for $\gamma_b/\gamma=10$, the infidelity $1-\fcal$ is tiny, being at least smaller than $10^{-2}$ for all the considered combinations of $\kappa$ and $G$, even when the adiabatic assumption $\gamma \ll \gamma_b$ is not fully respected. When the ratio is set to $\gamma_b=100\gamma$, we have $1-\fcal\ll 10^{-4}$ in all cases tested, indicating that the effective model we derived provides a very accurate description of the full model in the adiabatic limit.

\bibliography{bib}

\begin{thebibliography}{46}%
\makeatletter
\providecommand \@ifxundefined [1]{%
 \@ifx{#1\undefined}
}%
\providecommand \@ifnum [1]{%
 \ifnum #1\expandafter \@firstoftwo
 \else \expandafter \@secondoftwo
 \fi
}%
\providecommand \@ifx [1]{%
 \ifx #1\expandafter \@firstoftwo
 \else \expandafter \@secondoftwo
 \fi
}%
\providecommand \natexlab [1]{#1}%
\providecommand \enquote  [1]{``#1''}%
\providecommand \bibnamefont  [1]{#1}%
\providecommand \bibfnamefont [1]{#1}%
\providecommand \citenamefont [1]{#1}%
\providecommand \href@noop [0]{\@secondoftwo}%
\providecommand \href [0]{\begingroup \@sanitize@url \@href}%
\providecommand \@href[1]{\@@startlink{#1}\@@href}%
\providecommand \@@href[1]{\endgroup#1\@@endlink}%
\providecommand \@sanitize@url [0]{\catcode `\\12\catcode `\$12\catcode
  `\&12\catcode `\#12\catcode `\^12\catcode `\_12\catcode `\%12\relax}%
\providecommand \@@startlink[1]{}%
\providecommand \@@endlink[0]{}%
\providecommand \url  [0]{\begingroup\@sanitize@url \@url }%
\providecommand \@url [1]{\endgroup\@href {#1}{\urlprefix }}%
\providecommand \urlprefix  [0]{URL }%
\providecommand \Eprint [0]{\href }%
\providecommand \doibase [0]{http://dx.doi.org/}%
\providecommand \selectlanguage [0]{\@gobble}%
\providecommand \bibinfo  [0]{\@secondoftwo}%
\providecommand \bibfield  [0]{\@secondoftwo}%
\providecommand \translation [1]{[#1]}%
\providecommand \BibitemOpen [0]{}%
\providecommand \bibitemStop [0]{}%
\providecommand \bibitemNoStop [0]{.\EOS\space}%
\providecommand \EOS [0]{\spacefactor3000\relax}%
\providecommand \BibitemShut  [1]{\csname bibitem#1\endcsname}%
\let\auto@bib@innerbib\@empty
\bibitem [{\citenamefont {Ramirez}(1994)}]{ramirez1994strongly}%
  \BibitemOpen
  \bibfield  {author} {\bibinfo {author} {\bibfnamefont {A.}~\bibnamefont
  {Ramirez}},\ }\href {\doibase 10.1146/annurev.ms.24.080194.002321} {\bibfield
   {journal} {\bibinfo  {journal} {Annual Review of Materials Science}\
  }\textbf {\bibinfo {volume} {24}},\ \bibinfo {pages} {453} (\bibinfo {year}
  {1994})}\BibitemShut {NoStop}%
\bibitem [{\citenamefont {Moessner}\ and\ \citenamefont
  {Ramirez}(2006)}]{moessner2006geometrical}%
  \BibitemOpen
  \bibfield  {author} {\bibinfo {author} {\bibfnamefont {R.}~\bibnamefont
  {Moessner}}\ and\ \bibinfo {author} {\bibfnamefont {A.~P.}\ \bibnamefont
  {Ramirez}},\ }\href {\doibase 10.1063/1.2186278} {\bibfield  {journal}
  {\bibinfo  {journal} {Phys. Today}\ }\textbf {\bibinfo {volume} {59}},\
  \bibinfo {pages} {24} (\bibinfo {year} {2006})}\BibitemShut {NoStop}%
\bibitem [{\citenamefont {Chalker}\ \emph {et~al.}(1992)\citenamefont
  {Chalker}, \citenamefont {Holdsworth},\ and\ \citenamefont
  {Shender}}]{chalker1992hidden}%
  \BibitemOpen
  \bibfield  {author} {\bibinfo {author} {\bibfnamefont {J.~T.}\ \bibnamefont
  {Chalker}}, \bibinfo {author} {\bibfnamefont {P.~C.}\ \bibnamefont
  {Holdsworth}}, \ and\ \bibinfo {author} {\bibfnamefont {E.}~\bibnamefont
  {Shender}},\ }\href {\doibase 10.1103/PhysRevLett.68.855} {\bibfield
  {journal} {\bibinfo  {journal} {Physical review letters}\ }\textbf {\bibinfo
  {volume} {68}},\ \bibinfo {pages} {855} (\bibinfo {year} {1992})}\BibitemShut
  {NoStop}%
\bibitem [{\citenamefont {Si}\ and\ \citenamefont
  {Abrahams}(2008)}]{si2008strong}%
  \BibitemOpen
  \bibfield  {author} {\bibinfo {author} {\bibfnamefont {Q.}~\bibnamefont
  {Si}}\ and\ \bibinfo {author} {\bibfnamefont {E.}~\bibnamefont {Abrahams}},\
  }\href {\doibase 10.1103/PhysRevLett.101.076401} {\bibfield  {journal}
  {\bibinfo  {journal} {Physical Review Letters}\ }\textbf {\bibinfo {volume}
  {101}},\ \bibinfo {pages} {076401} (\bibinfo {year} {2008})}\BibitemShut
  {NoStop}%
\bibitem [{\citenamefont {Le~Hur}\ and\ \citenamefont {Rice}(2009)}]{lehur09}%
  \BibitemOpen
  \bibfield  {author} {\bibinfo {author} {\bibfnamefont {K.}~\bibnamefont
  {Le~Hur}}\ and\ \bibinfo {author} {\bibfnamefont {M.}~\bibnamefont {Rice}},\
  }\href {\doibase 10.1016/j.aop.2009.02.004} {\bibfield  {journal} {\bibinfo
  {journal} {Annals of Physics}\ }\textbf {\bibinfo {volume} {324}},\ \bibinfo
  {pages} {1452 } (\bibinfo {year} {2009})}\BibitemShut {NoStop}%
\bibitem [{\citenamefont {Ramires}(2019)}]{ramires19}%
  \BibitemOpen
  \bibfield  {author} {\bibinfo {author} {\bibfnamefont {A.}~\bibnamefont
  {Ramires}},\ }\href {\doibase 10.1038/s41567-019-0668-4} {\bibfield
  {journal} {\bibinfo  {journal} {Nature Physics}\ }\textbf {\bibinfo {volume}
  {15}},\ \bibinfo {pages} {1212} (\bibinfo {year} {2019})}\BibitemShut
  {NoStop}%
\bibitem [{\citenamefont {Giauque}\ and\ \citenamefont
  {Stout}(1936)}]{giauque1936entropy}%
  \BibitemOpen
  \bibfield  {author} {\bibinfo {author} {\bibfnamefont {W.}~\bibnamefont
  {Giauque}}\ and\ \bibinfo {author} {\bibfnamefont {J.}~\bibnamefont
  {Stout}},\ }\href {\doibase 10.1021/ja01298a023} {\bibfield  {journal}
  {\bibinfo  {journal} {Journal of the American Chemical Society}\ }\textbf
  {\bibinfo {volume} {58}},\ \bibinfo {pages} {1144} (\bibinfo {year}
  {1936})}\BibitemShut {NoStop}%
\bibitem [{\citenamefont {Wannier}(1950)}]{wannier1950antiferromagnetism}%
  \BibitemOpen
  \bibfield  {author} {\bibinfo {author} {\bibfnamefont {G.}~\bibnamefont
  {Wannier}},\ }\href {\doibase 10.1103/PhysRev.79.357} {\bibfield  {journal}
  {\bibinfo  {journal} {Physical Review}\ }\textbf {\bibinfo {volume} {79}},\
  \bibinfo {pages} {357} (\bibinfo {year} {1950})}\BibitemShut {NoStop}%
\bibitem [{\citenamefont {M{\'e}zard}\ \emph {et~al.}(1987)\citenamefont
  {M{\'e}zard}, \citenamefont {Parisi},\ and\ \citenamefont
  {Virasoro}}]{mezard1987spin}%
  \BibitemOpen
  \bibfield  {author} {\bibinfo {author} {\bibfnamefont {M.}~\bibnamefont
  {M{\'e}zard}}, \bibinfo {author} {\bibfnamefont {G.}~\bibnamefont {Parisi}},
  \ and\ \bibinfo {author} {\bibfnamefont {M.}~\bibnamefont {Virasoro}},\
  }\href {\doibase 10.1142/0271} {\emph {\bibinfo {title} {Spin glass theory
  and beyond: An Introduction to the Replica Method and Its Applications}}},\
  Vol.~\bibinfo {volume} {9}\ (\bibinfo  {publisher} {World Scientific
  Publishing Company},\ \bibinfo {year} {1987})\BibitemShut {NoStop}%
\bibitem [{\citenamefont {Balents}(2010)}]{balents2010spin}%
  \BibitemOpen
  \bibfield  {author} {\bibinfo {author} {\bibfnamefont {L.}~\bibnamefont
  {Balents}},\ }\href {\doibase 10.1038/nature08917} {\bibfield  {journal}
  {\bibinfo  {journal} {Nature}\ }\textbf {\bibinfo {volume} {464}},\ \bibinfo
  {pages} {199} (\bibinfo {year} {2010})}\BibitemShut {NoStop}%
\bibitem [{\citenamefont {Zhou}\ \emph {et~al.}(2017)\citenamefont {Zhou},
  \citenamefont {Kanoda},\ and\ \citenamefont {Ng}}]{zhou2017quantum}%
  \BibitemOpen
  \bibfield  {author} {\bibinfo {author} {\bibfnamefont {Y.}~\bibnamefont
  {Zhou}}, \bibinfo {author} {\bibfnamefont {K.}~\bibnamefont {Kanoda}}, \ and\
  \bibinfo {author} {\bibfnamefont {T.-K.}\ \bibnamefont {Ng}},\ }\href
  {\doibase 10.1103/RevModPhys.89.025003} {\bibfield  {journal} {\bibinfo
  {journal} {Reviews of Modern Physics}\ }\textbf {\bibinfo {volume} {89}},\
  \bibinfo {pages} {025003} (\bibinfo {year} {2017})}\BibitemShut {NoStop}%
\bibitem [{\citenamefont {Yan}\ \emph {et~al.}(2011)\citenamefont {Yan},
  \citenamefont {Huse},\ and\ \citenamefont {White}}]{yan2011spin}%
  \BibitemOpen
  \bibfield  {author} {\bibinfo {author} {\bibfnamefont {S.}~\bibnamefont
  {Yan}}, \bibinfo {author} {\bibfnamefont {D.~A.}\ \bibnamefont {Huse}}, \
  and\ \bibinfo {author} {\bibfnamefont {S.~R.}\ \bibnamefont {White}},\ }\href
  {\doibase 10.1126/science.1201080} {\bibfield  {journal} {\bibinfo  {journal}
  {Science}\ }\textbf {\bibinfo {volume} {332}},\ \bibinfo {pages} {1173}
  (\bibinfo {year} {2011})}\BibitemShut {NoStop}%
\bibitem [{\citenamefont {Shimizu}\ \emph {et~al.}(2003)\citenamefont
  {Shimizu}, \citenamefont {Miyagawa}, \citenamefont {Kanoda}, \citenamefont
  {Maesato},\ and\ \citenamefont {Saito}}]{shimizu2003spin}%
  \BibitemOpen
  \bibfield  {author} {\bibinfo {author} {\bibfnamefont {Y.}~\bibnamefont
  {Shimizu}}, \bibinfo {author} {\bibfnamefont {K.}~\bibnamefont {Miyagawa}},
  \bibinfo {author} {\bibfnamefont {K.}~\bibnamefont {Kanoda}}, \bibinfo
  {author} {\bibfnamefont {M.}~\bibnamefont {Maesato}}, \ and\ \bibinfo
  {author} {\bibfnamefont {G.}~\bibnamefont {Saito}},\ }\href {\doibase
  10.1103/PhysRevLett.91.107001} {\bibfield  {journal} {\bibinfo  {journal}
  {Physical review letters}\ }\textbf {\bibinfo {volume} {91}},\ \bibinfo
  {pages} {107001} (\bibinfo {year} {2003})}\BibitemShut {NoStop}%
\bibitem [{\citenamefont {Coldea}\ \emph {et~al.}(2001)\citenamefont {Coldea},
  \citenamefont {Tennant}, \citenamefont {Tsvelik},\ and\ \citenamefont
  {Tylczynski}}]{coldea2001experimental}%
  \BibitemOpen
  \bibfield  {author} {\bibinfo {author} {\bibfnamefont {R.}~\bibnamefont
  {Coldea}}, \bibinfo {author} {\bibfnamefont {D.}~\bibnamefont {Tennant}},
  \bibinfo {author} {\bibfnamefont {A.}~\bibnamefont {Tsvelik}}, \ and\
  \bibinfo {author} {\bibfnamefont {Z.}~\bibnamefont {Tylczynski}},\ }\href
  {\doibase 10.1103/PhysRevLett.86.1335} {\bibfield  {journal} {\bibinfo
  {journal} {Physical review letters}\ }\textbf {\bibinfo {volume} {86}},\
  \bibinfo {pages} {1335} (\bibinfo {year} {2001})}\BibitemShut {NoStop}%
\bibitem [{\citenamefont {Savary}\ and\ \citenamefont
  {Balents}(2016)}]{savary2016quantum}%
  \BibitemOpen
  \bibfield  {author} {\bibinfo {author} {\bibfnamefont {L.}~\bibnamefont
  {Savary}}\ and\ \bibinfo {author} {\bibfnamefont {L.}~\bibnamefont
  {Balents}},\ }\href {\doibase 10.1088/0034-4885/80/1/016502} {\bibfield
  {journal} {\bibinfo  {journal} {Reports on Progress in Physics}\ }\textbf
  {\bibinfo {volume} {80}},\ \bibinfo {pages} {016502} (\bibinfo {year}
  {2016})}\BibitemShut {NoStop}%
\bibitem [{\citenamefont {Weimer}\ \emph {et~al.}(2010)\citenamefont {Weimer},
  \citenamefont {Müller}, \citenamefont {Lesanovsky}, \citenamefont {Zoller},\
  and\ \citenamefont {Büchler}}]{buchler10}%
  \BibitemOpen
  \bibfield  {author} {\bibinfo {author} {\bibfnamefont {H.}~\bibnamefont
  {Weimer}}, \bibinfo {author} {\bibfnamefont {M.}~\bibnamefont {Müller}},
  \bibinfo {author} {\bibfnamefont {I.}~\bibnamefont {Lesanovsky}}, \bibinfo
  {author} {\bibfnamefont {P.}~\bibnamefont {Zoller}}, \ and\ \bibinfo {author}
  {\bibfnamefont {H.~P.}\ \bibnamefont {Büchler}},\ }\href {\doibase
  10.1038/nphys1614} {\bibfield  {journal} {\bibinfo  {journal} {Nature
  Physics}\ }\textbf {\bibinfo {volume} {6}},\ \bibinfo {pages} {382–388}
  (\bibinfo {year} {2010})}\BibitemShut {NoStop}%
\bibitem [{\citenamefont {Cantat-Moltrecht}\ \emph {et~al.}(2020)\citenamefont
  {Cantat-Moltrecht}, \citenamefont {Corti\~nas}, \citenamefont {Ravon},
  \citenamefont {M\'ehaignerie}, \citenamefont {Haroche}, \citenamefont
  {Raimond}, \citenamefont {Favier}, \citenamefont {Brune},\ and\ \citenamefont
  {Sayrin}}]{sayrin20}%
  \BibitemOpen
  \bibfield  {author} {\bibinfo {author} {\bibfnamefont {T.}~\bibnamefont
  {Cantat-Moltrecht}}, \bibinfo {author} {\bibfnamefont {R.}~\bibnamefont
  {Corti\~nas}}, \bibinfo {author} {\bibfnamefont {B.}~\bibnamefont {Ravon}},
  \bibinfo {author} {\bibfnamefont {P.}~\bibnamefont {M\'ehaignerie}}, \bibinfo
  {author} {\bibfnamefont {S.}~\bibnamefont {Haroche}}, \bibinfo {author}
  {\bibfnamefont {J.~M.}\ \bibnamefont {Raimond}}, \bibinfo {author}
  {\bibfnamefont {M.}~\bibnamefont {Favier}}, \bibinfo {author} {\bibfnamefont
  {M.}~\bibnamefont {Brune}}, \ and\ \bibinfo {author} {\bibfnamefont
  {C.}~\bibnamefont {Sayrin}},\ }\href {\doibase
  10.1103/PhysRevResearch.2.022032} {\bibfield  {journal} {\bibinfo  {journal}
  {Phys. Rev. Research}\ }\textbf {\bibinfo {volume} {2}},\ \bibinfo {pages}
  {022032} (\bibinfo {year} {2020})}\BibitemShut {NoStop}%
\bibitem [{\citenamefont {Kuhr}(2016)}]{kuhr16}%
  \BibitemOpen
  \bibfield  {author} {\bibinfo {author} {\bibfnamefont {S.}~\bibnamefont
  {Kuhr}},\ }\href {\doibase 10.1093/nsr/nww023} {\bibfield  {journal}
  {\bibinfo  {journal} {National Science Review}\ }\textbf {\bibinfo {volume}
  {3}},\ \bibinfo {pages} {170} (\bibinfo {year} {2016})}\BibitemShut {NoStop}%
\bibitem [{\citenamefont {Aspuru-Guzik}\ and\ \citenamefont
  {Walther}(2012)}]{aspuru2012photonic}%
  \BibitemOpen
  \bibfield  {author} {\bibinfo {author} {\bibfnamefont {A.}~\bibnamefont
  {Aspuru-Guzik}}\ and\ \bibinfo {author} {\bibfnamefont {P.}~\bibnamefont
  {Walther}},\ }\href {\doibase 10.1038/nphys2253} {\bibfield  {journal}
  {\bibinfo  {journal} {Nature physics}\ }\textbf {\bibinfo {volume} {8}},\
  \bibinfo {pages} {285} (\bibinfo {year} {2012})}\BibitemShut {NoStop}%
\bibitem [{\citenamefont {Hartmann}(2016)}]{hartmann2016quantum}%
  \BibitemOpen
  \bibfield  {author} {\bibinfo {author} {\bibfnamefont {M.~J.}\ \bibnamefont
  {Hartmann}},\ }\href {\doibase 10.1088/2040-8978/18/10/104005} {\bibfield
  {journal} {\bibinfo  {journal} {Journal of Optics}\ }\textbf {\bibinfo
  {volume} {18}},\ \bibinfo {pages} {104005} (\bibinfo {year}
  {2016})}\BibitemShut {NoStop}%
\bibitem [{\citenamefont {Angelakis}(2017)}]{angelakis2017quantum}%
  \BibitemOpen
  \bibfield  {author} {\bibinfo {author} {\bibfnamefont {D.~G.}\ \bibnamefont
  {Angelakis}},\ }\href {\doibase 10.1007/978-3-319-52025-4} {\emph {\bibinfo
  {title} {Quantum Simulations with Photons and Polaritons}}}\ (\bibinfo
  {publisher} {Springer},\ \bibinfo {year} {2017})\BibitemShut {NoStop}%
\bibitem [{\citenamefont {Carusotto}\ and\ \citenamefont
  {Ciuti}(2013)}]{carusotto2013quantum}%
  \BibitemOpen
  \bibfield  {author} {\bibinfo {author} {\bibfnamefont {I.}~\bibnamefont
  {Carusotto}}\ and\ \bibinfo {author} {\bibfnamefont {C.}~\bibnamefont
  {Ciuti}},\ }\href {\doibase 10.1103/RevModPhys.85.299} {\bibfield  {journal}
  {\bibinfo  {journal} {Reviews of Modern Physics}\ }\textbf {\bibinfo {volume}
  {85}},\ \bibinfo {pages} {299} (\bibinfo {year} {2013})}\BibitemShut
  {NoStop}%
\bibitem [{\citenamefont {Amo}\ and\ \citenamefont
  {Bloch}(2016)}]{amo2016exciton}%
  \BibitemOpen
  \bibfield  {author} {\bibinfo {author} {\bibfnamefont {A.}~\bibnamefont
  {Amo}}\ and\ \bibinfo {author} {\bibfnamefont {J.}~\bibnamefont {Bloch}},\
  }\href {\doibase 10.1016/j.crhy.2016.08.007} {\bibfield  {journal} {\bibinfo
  {journal} {Comptes Rendus Physique}\ }\textbf {\bibinfo {volume} {17}},\
  \bibinfo {pages} {934} (\bibinfo {year} {2016})}\BibitemShut {NoStop}%
\bibitem [{\citenamefont {Boulier}\ \emph {et~al.}(2020)\citenamefont
  {Boulier}, \citenamefont {Jacquet}, \citenamefont {Ma{\^\i}tre},
  \citenamefont {Lerario}, \citenamefont {Claude}, \citenamefont {Pigeon},
  \citenamefont {Glorieux}, \citenamefont {Amo}, \citenamefont {Bloch},
  \citenamefont {Bramati} \emph {et~al.}}]{boulier2020microcavity}%
  \BibitemOpen
  \bibfield  {author} {\bibinfo {author} {\bibfnamefont {T.}~\bibnamefont
  {Boulier}}, \bibinfo {author} {\bibfnamefont {M.~J.}\ \bibnamefont
  {Jacquet}}, \bibinfo {author} {\bibfnamefont {A.}~\bibnamefont
  {Ma{\^\i}tre}}, \bibinfo {author} {\bibfnamefont {G.}~\bibnamefont
  {Lerario}}, \bibinfo {author} {\bibfnamefont {F.}~\bibnamefont {Claude}},
  \bibinfo {author} {\bibfnamefont {S.}~\bibnamefont {Pigeon}}, \bibinfo
  {author} {\bibfnamefont {Q.}~\bibnamefont {Glorieux}}, \bibinfo {author}
  {\bibfnamefont {A.}~\bibnamefont {Amo}}, \bibinfo {author} {\bibfnamefont
  {J.}~\bibnamefont {Bloch}}, \bibinfo {author} {\bibfnamefont
  {A.}~\bibnamefont {Bramati}},  \emph {et~al.},\ }\href {\doibase
  0.1002/qute.202000052} {\bibfield  {journal} {\bibinfo  {journal} {Advanced
  Quantum Technologies}\ ,\ \bibinfo {pages} {2000052}} (\bibinfo {year}
  {2020})}\BibitemShut {NoStop}%
\bibitem [{\citenamefont {Berloff}\ \emph {et~al.}(2017)\citenamefont
  {Berloff}, \citenamefont {Silva}, \citenamefont {Kalinin}, \citenamefont
  {Askitopoulos}, \citenamefont {T{\"o}pfer}, \citenamefont {Cilibrizzi},
  \citenamefont {Langbein},\ and\ \citenamefont
  {Lagoudakis}}]{berloff2017realizing}%
  \BibitemOpen
  \bibfield  {author} {\bibinfo {author} {\bibfnamefont {N.~G.}\ \bibnamefont
  {Berloff}}, \bibinfo {author} {\bibfnamefont {M.}~\bibnamefont {Silva}},
  \bibinfo {author} {\bibfnamefont {K.}~\bibnamefont {Kalinin}}, \bibinfo
  {author} {\bibfnamefont {A.}~\bibnamefont {Askitopoulos}}, \bibinfo {author}
  {\bibfnamefont {J.~D.}\ \bibnamefont {T{\"o}pfer}}, \bibinfo {author}
  {\bibfnamefont {P.}~\bibnamefont {Cilibrizzi}}, \bibinfo {author}
  {\bibfnamefont {W.}~\bibnamefont {Langbein}}, \ and\ \bibinfo {author}
  {\bibfnamefont {P.~G.}\ \bibnamefont {Lagoudakis}},\ }\href {\doibase
  10.1038/nmat4971} {\bibfield  {journal} {\bibinfo  {journal} {Nature
  materials}\ }\textbf {\bibinfo {volume} {16}},\ \bibinfo {pages} {1120}
  (\bibinfo {year} {2017})}\BibitemShut {NoStop}%
\bibitem [{\citenamefont {Goblot}\ \emph {et~al.}(2019)\citenamefont {Goblot},
  \citenamefont {Rauer}, \citenamefont {Vicentini}, \citenamefont
  {Le~Boit{\'e}}, \citenamefont {Galopin}, \citenamefont {Lema{\^\i}tre},
  \citenamefont {Le~Gratiet}, \citenamefont {Harouri}, \citenamefont {Sagnes},
  \citenamefont {Ravets} \emph {et~al.}}]{goblot2019nonlinear}%
  \BibitemOpen
  \bibfield  {author} {\bibinfo {author} {\bibfnamefont {V.}~\bibnamefont
  {Goblot}}, \bibinfo {author} {\bibfnamefont {B.}~\bibnamefont {Rauer}},
  \bibinfo {author} {\bibfnamefont {F.}~\bibnamefont {Vicentini}}, \bibinfo
  {author} {\bibfnamefont {A.}~\bibnamefont {Le~Boit{\'e}}}, \bibinfo {author}
  {\bibfnamefont {E.}~\bibnamefont {Galopin}}, \bibinfo {author} {\bibfnamefont
  {A.}~\bibnamefont {Lema{\^\i}tre}}, \bibinfo {author} {\bibfnamefont
  {L.}~\bibnamefont {Le~Gratiet}}, \bibinfo {author} {\bibfnamefont
  {A.}~\bibnamefont {Harouri}}, \bibinfo {author} {\bibfnamefont
  {I.}~\bibnamefont {Sagnes}}, \bibinfo {author} {\bibfnamefont
  {S.}~\bibnamefont {Ravets}},  \emph {et~al.},\ }\href {\doibase
  10.1103/PhysRevLett.123.113901} {\bibfield  {journal} {\bibinfo  {journal}
  {Physical Review Letters}\ }\textbf {\bibinfo {volume} {123}},\ \bibinfo
  {pages} {113901} (\bibinfo {year} {2019})}\BibitemShut {NoStop}%
\bibitem [{\citenamefont {Haroche}\ \emph {et~al.}(2020)\citenamefont
  {Haroche}, \citenamefont {Brune},\ and\ \citenamefont {Raimond}}]{haroche20}%
  \BibitemOpen
  \bibfield  {author} {\bibinfo {author} {\bibfnamefont {S.}~\bibnamefont
  {Haroche}}, \bibinfo {author} {\bibfnamefont {M.}~\bibnamefont {Brune}}, \
  and\ \bibinfo {author} {\bibfnamefont {J.}~\bibnamefont {Raimond}},\ }\href
  {\doibase 10.1038/s41567-020-0812-1} {\bibfield  {journal} {\bibinfo
  {journal} {Nature Physics}\ }\textbf {\bibinfo {volume} {16}},\ \bibinfo
  {pages} {243} (\bibinfo {year} {2020})}\BibitemShut {NoStop}%
\bibitem [{\citenamefont {Schmidt}\ and\ \citenamefont
  {Koch}(2013)}]{schmidt2013circuit}%
  \BibitemOpen
  \bibfield  {author} {\bibinfo {author} {\bibfnamefont {S.}~\bibnamefont
  {Schmidt}}\ and\ \bibinfo {author} {\bibfnamefont {J.}~\bibnamefont {Koch}},\
  }\href {\doibase 10.1002/andp.201200261} {\bibfield  {journal} {\bibinfo
  {journal} {Annalen der Physik}\ }\textbf {\bibinfo {volume} {525}},\ \bibinfo
  {pages} {395} (\bibinfo {year} {2013})}\BibitemShut {NoStop}%
\bibitem [{noa(2020)}]{noauthor_abc_2020}%
  \BibitemOpen
  \href {\doibase 10.1038/s41567-020-0847-3} {\bibfield  {journal} {\bibinfo
  {journal} {Nature Physics}\ }\textbf {\bibinfo {volume} {16}},\ \bibinfo
  {pages} {233} (\bibinfo {year} {2020})}\BibitemShut {NoStop}%
\bibitem [{\citenamefont {Minganti}\ \emph {et~al.}(2016)\citenamefont
  {Minganti}, \citenamefont {Bartolo}, \citenamefont {Lolli}, \citenamefont
  {Casteels},\ and\ \citenamefont {Ciuti}}]{minganti2016exact}%
  \BibitemOpen
  \bibfield  {author} {\bibinfo {author} {\bibfnamefont {F.}~\bibnamefont
  {Minganti}}, \bibinfo {author} {\bibfnamefont {N.}~\bibnamefont {Bartolo}},
  \bibinfo {author} {\bibfnamefont {J.}~\bibnamefont {Lolli}}, \bibinfo
  {author} {\bibfnamefont {W.}~\bibnamefont {Casteels}}, \ and\ \bibinfo
  {author} {\bibfnamefont {C.}~\bibnamefont {Ciuti}},\ }\href {\doibase
  10.1038/srep26987} {\bibfield  {journal} {\bibinfo  {journal} {Scientific
  reports}\ }\textbf {\bibinfo {volume} {6}},\ \bibinfo {pages} {1} (\bibinfo
  {year} {2016})}\BibitemShut {NoStop}%
\bibitem [{\citenamefont {Bartolo}\ \emph {et~al.}(2016)\citenamefont
  {Bartolo}, \citenamefont {Minganti}, \citenamefont {Casteels},\ and\
  \citenamefont {Ciuti}}]{bartolo2016exact}%
  \BibitemOpen
  \bibfield  {author} {\bibinfo {author} {\bibfnamefont {N.}~\bibnamefont
  {Bartolo}}, \bibinfo {author} {\bibfnamefont {F.}~\bibnamefont {Minganti}},
  \bibinfo {author} {\bibfnamefont {W.}~\bibnamefont {Casteels}}, \ and\
  \bibinfo {author} {\bibfnamefont {C.}~\bibnamefont {Ciuti}},\ }\href
  {\doibase 10.1103/PhysRevA.94.033841} {\bibfield  {journal} {\bibinfo
  {journal} {Physical Review A}\ }\textbf {\bibinfo {volume} {94}},\ \bibinfo
  {pages} {033841} (\bibinfo {year} {2016})}\BibitemShut {NoStop}%
\bibitem [{\citenamefont {Rota}\ \emph {et~al.}(2019)\citenamefont {Rota},
  \citenamefont {Minganti}, \citenamefont {Ciuti},\ and\ \citenamefont
  {Savona}}]{rota2019quantum}%
  \BibitemOpen
  \bibfield  {author} {\bibinfo {author} {\bibfnamefont {R.}~\bibnamefont
  {Rota}}, \bibinfo {author} {\bibfnamefont {F.}~\bibnamefont {Minganti}},
  \bibinfo {author} {\bibfnamefont {C.}~\bibnamefont {Ciuti}}, \ and\ \bibinfo
  {author} {\bibfnamefont {V.}~\bibnamefont {Savona}},\ }\href {\doibase
  10.1103/PhysRevLett.122.110405} {\bibfield  {journal} {\bibinfo  {journal}
  {Physical Review Letters}\ }\textbf {\bibinfo {volume} {122}},\ \bibinfo
  {pages} {110405} (\bibinfo {year} {2019})}\BibitemShut {NoStop}%
\bibitem [{\citenamefont {Bartolo}\ \emph {et~al.}(2017)\citenamefont
  {Bartolo}, \citenamefont {Minganti}, \citenamefont {Lolli},\ and\
  \citenamefont {Ciuti}}]{bartolo2017homodyne}%
  \BibitemOpen
  \bibfield  {author} {\bibinfo {author} {\bibfnamefont {N.}~\bibnamefont
  {Bartolo}}, \bibinfo {author} {\bibfnamefont {F.}~\bibnamefont {Minganti}},
  \bibinfo {author} {\bibfnamefont {J.}~\bibnamefont {Lolli}}, \ and\ \bibinfo
  {author} {\bibfnamefont {C.}~\bibnamefont {Ciuti}},\ }\href {\doibase
  10.1140/epjst/e2016-60385-8} {\bibfield  {journal} {\bibinfo  {journal} {The
  European Physical Journal Special Topics}\ }\textbf {\bibinfo {volume}
  {226}},\ \bibinfo {pages} {2705} (\bibinfo {year} {2017})}\BibitemShut
  {NoStop}%
\bibitem [{\citenamefont {Mirrahimi}\ \emph {et~al.}(2014)\citenamefont
  {Mirrahimi}, \citenamefont {Leghtas}, \citenamefont {Albert}, \citenamefont
  {Touzard}, \citenamefont {Schoelkopf}, \citenamefont {Jiang},\ and\
  \citenamefont {Devoret}}]{mirrahimi2014dynamically}%
  \BibitemOpen
  \bibfield  {author} {\bibinfo {author} {\bibfnamefont {M.}~\bibnamefont
  {Mirrahimi}}, \bibinfo {author} {\bibfnamefont {Z.}~\bibnamefont {Leghtas}},
  \bibinfo {author} {\bibfnamefont {V.~V.}\ \bibnamefont {Albert}}, \bibinfo
  {author} {\bibfnamefont {S.}~\bibnamefont {Touzard}}, \bibinfo {author}
  {\bibfnamefont {R.~J.}\ \bibnamefont {Schoelkopf}}, \bibinfo {author}
  {\bibfnamefont {L.}~\bibnamefont {Jiang}}, \ and\ \bibinfo {author}
  {\bibfnamefont {M.~H.}\ \bibnamefont {Devoret}},\ }\href {\doibase
  10.1088/1367-2630/16/4/045014} {\bibfield  {journal} {\bibinfo  {journal}
  {New Journal of Physics}\ }\textbf {\bibinfo {volume} {16}},\ \bibinfo
  {pages} {045014} (\bibinfo {year} {2014})}\BibitemShut {NoStop}%
\bibitem [{\citenamefont {Leghtas}\ \emph {et~al.}(2015)\citenamefont
  {Leghtas}, \citenamefont {Touzard}, \citenamefont {Pop}, \citenamefont {Kou},
  \citenamefont {Vlastakis}, \citenamefont {Petrenko}, \citenamefont {Sliwa},
  \citenamefont {Narla}, \citenamefont {Shankar}, \citenamefont {Hatridge}
  \emph {et~al.}}]{leghtas2015confining}%
  \BibitemOpen
  \bibfield  {author} {\bibinfo {author} {\bibfnamefont {Z.}~\bibnamefont
  {Leghtas}}, \bibinfo {author} {\bibfnamefont {S.}~\bibnamefont {Touzard}},
  \bibinfo {author} {\bibfnamefont {I.~M.}\ \bibnamefont {Pop}}, \bibinfo
  {author} {\bibfnamefont {A.}~\bibnamefont {Kou}}, \bibinfo {author}
  {\bibfnamefont {B.}~\bibnamefont {Vlastakis}}, \bibinfo {author}
  {\bibfnamefont {A.}~\bibnamefont {Petrenko}}, \bibinfo {author}
  {\bibfnamefont {K.~M.}\ \bibnamefont {Sliwa}}, \bibinfo {author}
  {\bibfnamefont {A.}~\bibnamefont {Narla}}, \bibinfo {author} {\bibfnamefont
  {S.}~\bibnamefont {Shankar}}, \bibinfo {author} {\bibfnamefont {M.~J.}\
  \bibnamefont {Hatridge}},  \emph {et~al.},\ }\href {\doibase
  10.1126/science.aaa2085} {\bibfield  {journal} {\bibinfo  {journal}
  {Science}\ }\textbf {\bibinfo {volume} {347}},\ \bibinfo {pages} {853}
  (\bibinfo {year} {2015})}\BibitemShut {NoStop}%
\bibitem [{\citenamefont {Devoret}\ and\ \citenamefont
  {Schoelkopf}(2013)}]{devoret2013superconducting}%
  \BibitemOpen
  \bibfield  {author} {\bibinfo {author} {\bibfnamefont {M.~H.}\ \bibnamefont
  {Devoret}}\ and\ \bibinfo {author} {\bibfnamefont {R.~J.}\ \bibnamefont
  {Schoelkopf}},\ }\href {\doibase 10.1126/science.1231930} {\bibfield
  {journal} {\bibinfo  {journal} {Science}\ }\textbf {\bibinfo {volume}
  {339}},\ \bibinfo {pages} {1169} (\bibinfo {year} {2013})}\BibitemShut
  {NoStop}%
\bibitem [{\citenamefont {Blais}\ \emph {et~al.}(2007)\citenamefont {Blais},
  \citenamefont {Gambetta}, \citenamefont {Wallraff}, \citenamefont {Schuster},
  \citenamefont {Girvin}, \citenamefont {Devoret},\ and\ \citenamefont
  {Schoelkopf}}]{blais2007quantum}%
  \BibitemOpen
  \bibfield  {author} {\bibinfo {author} {\bibfnamefont {A.}~\bibnamefont
  {Blais}}, \bibinfo {author} {\bibfnamefont {J.}~\bibnamefont {Gambetta}},
  \bibinfo {author} {\bibfnamefont {A.}~\bibnamefont {Wallraff}}, \bibinfo
  {author} {\bibfnamefont {D.~I.}\ \bibnamefont {Schuster}}, \bibinfo {author}
  {\bibfnamefont {S.~M.}\ \bibnamefont {Girvin}}, \bibinfo {author}
  {\bibfnamefont {M.~H.}\ \bibnamefont {Devoret}}, \ and\ \bibinfo {author}
  {\bibfnamefont {R.~J.}\ \bibnamefont {Schoelkopf}},\ }\href {\doibase
  10.1103/PhysRevA.75.032329} {\bibfield  {journal} {\bibinfo  {journal}
  {Physical Review A}\ }\textbf {\bibinfo {volume} {75}},\ \bibinfo {pages}
  {032329} (\bibinfo {year} {2007})}\BibitemShut {NoStop}%
\bibitem [{\citenamefont {Schoelkopf}\ and\ \citenamefont
  {Girvin}(2008)}]{schoelkopf2008wiring}%
  \BibitemOpen
  \bibfield  {author} {\bibinfo {author} {\bibfnamefont {R.}~\bibnamefont
  {Schoelkopf}}\ and\ \bibinfo {author} {\bibfnamefont {S.}~\bibnamefont
  {Girvin}},\ }\href {\doibase 10.1038/451664a} {\bibfield  {journal} {\bibinfo
   {journal} {Nature}\ }\textbf {\bibinfo {volume} {451}},\ \bibinfo {pages}
  {664} (\bibinfo {year} {2008})}\BibitemShut {NoStop}%
\bibitem [{\citenamefont {Tsomokos}\ \emph {et~al.}(2010)\citenamefont
  {Tsomokos}, \citenamefont {Ashhab},\ and\ \citenamefont
  {Nori}}]{tsomokos2010using}%
  \BibitemOpen
  \bibfield  {author} {\bibinfo {author} {\bibfnamefont {D.~I.}\ \bibnamefont
  {Tsomokos}}, \bibinfo {author} {\bibfnamefont {S.}~\bibnamefont {Ashhab}}, \
  and\ \bibinfo {author} {\bibfnamefont {F.}~\bibnamefont {Nori}},\ }\href
  {\doibase 10.1103/PhysRevA.82.052311} {\bibfield  {journal} {\bibinfo
  {journal} {Physical Review A}\ }\textbf {\bibinfo {volume} {82}},\ \bibinfo
  {pages} {052311} (\bibinfo {year} {2010})}\BibitemShut {NoStop}%
\bibitem [{\citenamefont {Houck}\ \emph {et~al.}(2012)\citenamefont {Houck},
  \citenamefont {T{\"u}reci},\ and\ \citenamefont {Koch}}]{houck2012chip}%
  \BibitemOpen
  \bibfield  {author} {\bibinfo {author} {\bibfnamefont {A.~A.}\ \bibnamefont
  {Houck}}, \bibinfo {author} {\bibfnamefont {H.~E.}\ \bibnamefont
  {T{\"u}reci}}, \ and\ \bibinfo {author} {\bibfnamefont {J.}~\bibnamefont
  {Koch}},\ }\href {\doibase 10.1038/nphys2251} {\bibfield  {journal} {\bibinfo
   {journal} {Nature Physics}\ }\textbf {\bibinfo {volume} {8}},\ \bibinfo
  {pages} {292} (\bibinfo {year} {2012})}\BibitemShut {NoStop}%
\bibitem [{\citenamefont {Rota}\ and\ \citenamefont
  {Savona}(2019)}]{rota2019simulating}%
  \BibitemOpen
  \bibfield  {author} {\bibinfo {author} {\bibfnamefont {R.}~\bibnamefont
  {Rota}}\ and\ \bibinfo {author} {\bibfnamefont {V.}~\bibnamefont {Savona}},\
  }\href {\doibase 10.1103/PhysRevA.100.013838} {\bibfield  {journal} {\bibinfo
   {journal} {Physical Review A}\ }\textbf {\bibinfo {volume} {100}},\ \bibinfo
  {pages} {013838} (\bibinfo {year} {2019})}\BibitemShut {NoStop}%
\bibitem [{\citenamefont {Kounalakis}\ \emph {et~al.}(2018)\citenamefont
  {Kounalakis}, \citenamefont {Dickel}, \citenamefont {Bruno}, \citenamefont
  {Langford},\ and\ \citenamefont {Steele}}]{kounalakis2018tuneable}%
  \BibitemOpen
  \bibfield  {author} {\bibinfo {author} {\bibfnamefont {M.}~\bibnamefont
  {Kounalakis}}, \bibinfo {author} {\bibfnamefont {C.}~\bibnamefont {Dickel}},
  \bibinfo {author} {\bibfnamefont {A.}~\bibnamefont {Bruno}}, \bibinfo
  {author} {\bibfnamefont {N.~K.}\ \bibnamefont {Langford}}, \ and\ \bibinfo
  {author} {\bibfnamefont {G.~A.}\ \bibnamefont {Steele}},\ }\href {\doibase
  10.1038/s41534-018-0088-9} {\bibfield  {journal} {\bibinfo  {journal} {npj
  Quantum Information}\ }\textbf {\bibinfo {volume} {4}},\ \bibinfo {pages} {1}
  (\bibinfo {year} {2018})}\BibitemShut {NoStop}%
\bibitem [{\citenamefont {Haddadi}\ \emph {et~al.}(2014)\citenamefont
  {Haddadi}, \citenamefont {Hamel}, \citenamefont {Beaudoin}, \citenamefont
  {Sagnes}, \citenamefont {Sauvan}, \citenamefont {Lalanne}, \citenamefont
  {Levenson},\ and\ \citenamefont {Yacomotti}}]{haddadi2014photonic}%
  \BibitemOpen
  \bibfield  {author} {\bibinfo {author} {\bibfnamefont {S.}~\bibnamefont
  {Haddadi}}, \bibinfo {author} {\bibfnamefont {P.}~\bibnamefont {Hamel}},
  \bibinfo {author} {\bibfnamefont {G.}~\bibnamefont {Beaudoin}}, \bibinfo
  {author} {\bibfnamefont {I.}~\bibnamefont {Sagnes}}, \bibinfo {author}
  {\bibfnamefont {C.}~\bibnamefont {Sauvan}}, \bibinfo {author} {\bibfnamefont
  {P.}~\bibnamefont {Lalanne}}, \bibinfo {author} {\bibfnamefont {J.~A.}\
  \bibnamefont {Levenson}}, \ and\ \bibinfo {author} {\bibfnamefont
  {A.}~\bibnamefont {Yacomotti}},\ }\href {\doibase 10.1364/OE.22.012359}
  {\bibfield  {journal} {\bibinfo  {journal} {Optics Express}\ }\textbf
  {\bibinfo {volume} {22}},\ \bibinfo {pages} {12359} (\bibinfo {year}
  {2014})}\BibitemShut {NoStop}%
\bibitem [{Note1()}]{Note1}%
  \BibitemOpen
  \bibinfo {note} {In Ref. \cite {rota2019quantum}, the photonic Hamiltonian
  was actually mapped into a Heisenberg $XY$ Hamiltonian by identifying the
  Schroedinger cat states $\propto (\vert \alpha \rangle \pm \vert - \alpha
  \rangle )$ with the spin states along the $z$-direction. However, in the
  regime of $\vert \alpha \vert \gg 1$, only the $\protect \hat {\sigma
  }^x_j\protect \hat {\sigma }^x_{j+1}$ terms survive in the $XY$ Hamiltonian
  and $\mathinner {|{\pm \alpha }\rangle }$ are asymptotically the $\protect
  \hat {\sigma }^x$ eigenstates. Therefore, we can identify it with an Ising
  model after a rotation of basis $\protect \hat {\sigma }^x\rightarrow
  \protect \hat {\sigma }^z$.}\BibitemShut {Stop}%
\bibitem [{\citenamefont {Stephenson}(1964)}]{stephenson1964ising}%
  \BibitemOpen
  \bibfield  {author} {\bibinfo {author} {\bibfnamefont {J.}~\bibnamefont
  {Stephenson}},\ }\href {\doibase 10.1063/1.1704202} {\bibfield  {journal}
  {\bibinfo  {journal} {Journal of Mathematical Physics}\ }\textbf {\bibinfo
  {volume} {5}},\ \bibinfo {pages} {1009} (\bibinfo {year} {1964})}\BibitemShut
  {NoStop}%
\bibitem [{Note2()}]{Note2}%
  \BibitemOpen
  \bibinfo {note} {Note that in our effective model we always have $\kappa >0$.
  The case where $\kappa =0$ is simulated only for illustrative purpose to show
  the direct effect of the dissipative coupling.}\BibitemShut {Stop}%
\end{thebibliography}%
\bibliographystyle{apsrev4-1}
\clearpage

\end{document}